\pdfoutput=0
\documentclass[11pt]{article}
\usepackage{epsfig}
\usepackage{amsfonts}
\usepackage{amsmath}
\usepackage{bm,bbm}
\usepackage{cite}
\usepackage{cancel}
\usepackage[hidelinks]{hyperref}

\allowdisplaybreaks[1]

\newcommand{\la}[1]{\label{#1}}
\newcommand{\be}{\begin{equation}}
\newcommand{\ee}{\end{equation}}
\newcommand{\ba}{\begin{eqnarray}}
\newcommand{\ea}{\end{eqnarray}}
\newcommand{\fig}{Fig.~}
\newcommand{\figs}{Figs.~}
\newcommand{\eq}{Eq.~}
\newcommand{\eqs}{Eqs.~}
\newcommand{\se}{Sec.~}
\newcommand{\ses}{Secs.~}
\newcommand{\nr}[1]{(\ref{#1})}
\newcommand{\tr}{{\rm tr\,}} % {{\rm Tr\,}}
\newcommand{\nn}{\nonumber \\}

\renewcommand{\vec}[1]{{\bf #1}}

\newcommand{\tfr}[2]{{\textstyle \frac{#1}{#2}\,}}

\renewcommand{\eq}{eq.~}
\renewcommand{\eqs}{eqs.~}
\renewcommand{\se}{sec.~}
\renewcommand{\ses}{secs.~}
\renewcommand{\fig}{fig.~}
\renewcommand{\figs}{figs.~}
 
\newcommand{\alphas}{\alpha_{\rm s}}
\newcommand{\Nf}{N_{\rm f}}
\newcommand{\Nc}{N_{\rm c}}

\newcommand{\rmO}{{\mathcal{O}}}

\def\lsi{\raise0.3ex\hbox{$<$\kern-0.75em\raise-1.1ex\hbox{$\sim$}}}
\def\gsi{\raise0.3ex\hbox{$>$\kern-0.75em\raise-1.1ex\hbox{$\sim$}}}
\newcommand{\lsim}{\mathop{\lsi}}

\newcommand{\nB}{n_\rmii{B}}

\newcommand{\rmi}[1]{{\mbox{\scriptsize #1}}}
\newcommand{\rmii}[1]{{\mbox{\tiny\rm{#1}}}}
\newcommand{\rmiii}[1]{{\mbox{\tiny{$\scriptstyle{\rm#1}$}}}}

\newcommand{\im}{\mathop{\mbox{Im}}}
\newcommand{\Tint}[1]{{\hbox{$\sum$}\!\!\!\!\!\!\!\int\,}_{\!\!\!\!\raise-0.9ex\hbox{$\scriptstyle{#1}$}}}
\newcommand{\Tinti}[1]{{{\Sigma}\!\!\!\!\raise0.3ex\hbox{$\int$}_\rmii{${#1}$}}}
\newcommand{\bi}{\begin{itemize}}
\newcommand{\ei}{\end{itemize}}
\newcommand{\hide}[1]{ }
\newcommand{\deltabar}{\raise-0.02em\hbox{$\bar{}$}\hspace*{-0.8mm}{\delta}}
\newcommand{\ddeltabar}{\raise-0.18em\hbox{$\bar{}$}\hspace*{-0.8mm}{\delta}}
\newcommand{\iB}{\rmii{B}}

\renewcommand{\P}{\mathcal{P}}

\newcommand{\Q}{\mathcal{Q}}

\newcommand{\R}{\mathcal{R}}

\newcommand{\V}{\mathcal{V}}

\newcommand{\F}{\mathcal{F}}
\newcommand{\G}{\mathcal{G}}
\newcommand{\E}{\mathcal{S}} % {\rmii{$E$}}
\newcommand{\N}{\mathcal{N}}

\newcommand{\iT}{\rmii{$T$}}
\newcommand{\der}{,}
\newcommand{\now}{\rmii{$0$}}
\newcommand{\field}{\widehat{\mathcal{Q}}}
\newcommand{\mpl}{m_\rmii{pl}} % {m_\rmii{Pl}}
\makeatletter \@addtoreset{equation}{section} \makeatother
\renewcommand{\theequation}{\arabic{section}.\arabic{equation}}
\makeatletter
\renewcommand\section{\@startsection {section}{1}{\z@}%
                                   {-5.5ex \@plus -1ex \@minus -.2ex}% bfr-
                                   {2.3ex \@plus.2ex}%
                                   {\normalfont\large\bfseries}}
\renewcommand\subsection{\@startsection{subsection}{2}{\z@}%
                                     {-3.25ex\@plus -1ex \@minus -.2ex}%
                                     {1.5ex \@plus .2ex}%
                                     {\normalfont\normalsize\bfseries}}
\renewcommand\thesection {\@arabic\c@section}
\renewcommand\thesubsection   {\thesection.\@arabic\c@subsection}
\renewcommand{\@seccntformat}[1]{%
\csname the#1\endcsname.\hspace{1.0em}}
\makeatother
\begin{document}

\flushbottom

\begin{titlepage}

\begin{flushright}
% OUTLINE  \\ 
% DRAFT \\ 
% NOTES \\ 
% arXiv:2507.12849 \\ 
December 2025
\end{flushright}
\begin{centering}
\vfill

{\Large{\bf
    Evolution of coupled scalar perturbations through \\[3mm]
    ~smooth reheating. 
    II.~Thermal fluctuation regime
}} 

\vspace{0.8cm}

M.~Laine\hspace*{0.4mm}$^\rmi{a}_{ }$,
S.~Procacci\hspace*{0.4mm}$^\rmi{b}_{ }$,
A.~Rogelj\hspace*{0.4mm}$^\rmi{a}_{ }$

\vspace{0.8cm}

$^\rmi{a}_{ }${\em
AEC, 
Institute for Theoretical Physics, 
University of Bern, \\ 
Sidlerstrasse 5, CH-3012 Bern, Switzerland \\}

\vspace*{0.3cm}

$^\rmi{b}_{ }${\em
D\'epartement de Physique Th\'eorique, 
% and Center for Astroparticle Physics,
Universit\'e de Gen\`eve, \\ 
24 quai Ernest Ansermet, 
CH-1211 Gen\`eve 4, Switzerland \\}

\vspace*{0.8cm}

\mbox{\bf Abstract}
 
\end{centering}

\vspace*{0.3cm}
 
\noindent
Curvature perturbations with short wavelengths exit 
the Hubble horizon when the universe may contain a thermal plasma  
in addition to an inflaton field that drives its expansion. 
We solve the corresponding fluctuation-dissipation 
dynamics at linear order, building upon a previously
established set of gauge-invariant evolution equations. 
The properties of the noise autocorrelator are constrained 
via a matching of equilibrium correlators 
to quantum-statistical physics deep inside the Hubble horizon.
The curvature power spectrum is determined numerically,
without slow-roll approximations or assumptions
about the equilibration of the inflaton field. 
As applications, we scrutinize two issues from recent literature: 
the model dependence of the thermally modified power spectrum 
as a function of freeze-out parameters, 
and the viability of embedding
warm inflation within the Standard Model
(we offer support for the latter proposal). 
The role of pre-horizon-exit acoustic oscillations is illustrated.

\vspace*{0.5cm}

\noindent
{\small
{\em Emails}: 
{\tt laine@itp.unibe.ch},
{\tt simona.procacci@unige.ch}, 
{\tt alica.rogelj@unibe.ch}
}

\vfill

%% %\noindent
%% %PACS numbers: 
%% %11.10.Wx, %        Finite temperature field theory
%% %11.15.Ha, %        Lattice gauge theory  
%% %12.38.Bx, %        Perturbative calculations in QCD
%% %12.38.Mh, %        Quark--gluon plasma
%% %14.40.Nd, %        Bottom mesons
%% %\\
%% %Keywords: cosmological perturbation theory [100] 
%%            inflation [100]
%%            particle physics - cosmology connection [50] 
%%            axions [50]

\end{titlepage}

\tableofcontents

%%%%%%%%%%%%%%%%%%%%%%%%%%% SECTION %%%%%%%%%%%%%%%%%%%%%%%%%%%%%%%%%%%%%%
%
\section{Introduction}
\la{se:intro}

New probes of the early universe can arise from precision
measurements at larger momenta (smaller distance scales) than has been 
the norm in the past. Already now, the accuracy of 
the Planck CMB data~\cite{planck6} has been surpassed at 
angular momenta $\ell > 2000$ by new generations of 
terrestrial observatories~\cite{act}, yielding improved constraints
on the primordial power spectrum. Further data
is arising from large-scale structure, 
notably with the forthcoming Euclid mission~\cite{euclid}. 
Information about the earliest
halos~\cite{jwst} may require larger initial inhomogeneities
than a straightforward extrapolation from the CMB domain
would suggest. Future studies of gravitational waves
will extend the sensitivity up to very high frequencies~\cite{uhf}. 

Within the inflationary paradigm, a given comoving momentum 
mode exits the Hubble horizon at an early time, and 
is then ``frozen'' for a long while, being insensitive to the 
physics that may be taking place at smaller distances. 
In a late universe, the mode re-enters inside the
Hubble horizon, and is again susceptible to causal phenomena. 
Increasing the comoving momentum, a mode exits later, and
re-enters earlier. Large momenta are therefore more sensitive
to the physics that separates inflation and the radiation-dominated
universe, referred to as the period of {\em reheating}. 

It is fair to say that the nature of reheating is not 
well understood at the moment. One reason is that it is model dependent, 
with its features varying according to the couplings that
connect the inflaton
field to the Standard Model (or its extension). 
Another reason is that, even
for known couplings, reheating represents a complicated process, given that
the redistribution of the inflating energy density into 
an equipartitioned plasma of light particles 
requires multiple hard and soft interactions. 

In the present paper, our goal is to address the physics 
of the reheating period in a somewhat model-independent way, 
by describing the inflaton-plasma interaction through 
an effective coupling of a low-energy description, 
and varying it in a broad range~\cite{alica}. 
We treat the system as being composed of two subsystems: 
an inflaton field ($\varphi$), which interacts weakly with 
a strongly self-interacting radiation plasma, 
the latter carrying a seed temperature~($T$). 
The coupling between 
$\varphi$ and the plasma is parameterized by
the function
$\Upsilon = \Upsilon(\varphi,T)$, which we refer to 
as the {\em inflaton equilibration rate}. Despite its simplicity, 
this system displays rich physics, manifested by the spectrum of
its density perturbations. However, the general features
of the dynamics can be described by a transparent set of 
equations, even if the nature of the solution depends on the
functional forms of the inflaton potential, and of $\Upsilon$.

Though our fundamental interest concerns the reheating
period, the physical system just outlined
coincides with what has been studied for the model 
of {\em warm inflation}~\cite{warm1,warm2}. The investigation 
of the corresponding density perturbations 
has a long history~\cite{R_1,R_2,R_3,R_4}, 
and insights from complementary
viewpoints~\cite{warm0,fixed_pt,therm}.
In the past years, 
apart from model building, there have been
new looks at the basic equations 
and their solutions~\cite{mehrdad,garcia,ballesteros,freese,mm,ramos,hook}, 
sometimes finding results that do not agree with previous estimates. 
Therefore, it appears worthwhile to test 
our reheating framework in the warm inflation context. 

In fact, 
it turned out that our setup goes beyond the approximations
often adopted for warm inflation in several respects.
Denoting by $k$ a comoving momentum, 
and by $a$ the cosmological scale factor, the key ingredients
of our study are:  

\bi

\item
We determine the thermal noise autocorrelator
so that it interpolates smoothly between 
the ``classical'' ($k/a \ll T$) and 
the ``quantum'' ($k/a \gg T$) domains, 
and is free from double counting (cf.\ \se\ref{se:noise}). 
In the same spirit, we impose
quantum-mechanical initial conditions, so that they 
determine the solution if noise and dissipation are 
absent, and otherwise get gradually damped
(cf.\ \se\ref{ss:initial}).

\item
We invoke no slow-roll approximation, neither for the background 
evolution, nor for perturbations. In other words, our 
equations incorporate the decoupling of the inflaton field, 
when its energy density is below that of radiation, and the presence 
of acoustic oscillations, when a plasma 
mode is inside the Hubble horizon. 

\item
We invoke no gauge fixing, 
neither for inflaton ($\delta\varphi$), 
nor plasma ($\delta T$, $\delta v$), 
nor metric ($h^{ }_0$, $h^{ }_\rmii{D}$, $h$, $\vartheta$)
perturbations, but rather solve a closed set
of equations for three gauge-invariant
curvature perturbations ($\R^{ }_\varphi$, 
$\R^{ }_v$, $\R^{ }_\iT$). 
As a nice check of our numerics, 
the values of $\R^{ }_\varphi$, 
$\R^{ }_v$ and $\R^{ }_\iT$
coincide while a mode is outside of the Hubble horizon. 

\item
We make no assumption about whether the inflaton field has 
equilibrated, but rather let our equations drive $\varphi$ dynamically
towards equilibrium
(via dissipation and fluctuations). 
It is 
assumed, however, that the plasma with which $\varphi$
interacts, is in local kinetic 
and helicity equilibrium (for recent discussions about these 
non-trivial assumptions, 
see refs.~\cite{therm,hook}). Our setup allows for  
the self-interactions of the plasma
to be strong; strongly interacting plasmas might 
open interesting avenues~\cite{helena}. 

\item
We treat $Q \equiv \Upsilon / (3 H)$ as a time-dependent variable, 
rather than fixing it to a constant while perturbations
evolve (here $H \equiv \dot{a}/a$ is the Hubble rate). 
The functional form of~$\Upsilon$ 
is determined by the underlying microscopic theory, and
by the background evolution of 
the inflaton field and the temperature. 

\ei
For these reasons, it is not guaranteed {\em a priori} whether 
our results agree exactly 
with the warm inflation ones. 
It is also not easy to anticipate 
the source of the differences, 
given that there are many points.
We stress, however, that the differences are not 
associated with numerical issues or model dependence, but 
rather with the model-independent foundations of non-equilibrium
quantum statistical physics in an expanding background. 

Our presentation is organized as follows. 
The basic equations, initial conditions, and approximations 
valid in certain limits, 
are reviewed in \se\ref{se:eqs}. Then we turn to
the main theoretical ingredient of this work, the determination of 
the noise autocorrelator, in \se\ref{se:noise}. The methods that we have
adopted for the numerical solution of the
equations from \se\ref{se:eqs} are discussed in \se\ref{se:numerics}, 
with illustrative benchmarks 
subsequently shown in \se\ref{se:benchmarks}.
A broad range of friction coefficients, 
interpolating between thermal and vacuum-like ones, 
are scanned in \se\ref{se:scans}, allowing us to compare with 
literature and scrutinize the universality of 
freeze-out predictions.  
The second application 
can be found in \se\ref{se:sm}, investigating a 
Standard Model embedded warm inflation scenario. 
Conclusions and an outlook are offered in \se\ref{se:concl}.

%%%%%%%%%%%%%%%%%%%%%%%%%%% SECTION %%%%%%%%%%%%%%%%%%%%%%%%%%%%%%%%%%%%%%
%
\section{Evolution equations and initial conditions for perturbations}
\la{se:eqs}

%%%%%%%%%%%%%%%%%%%%%%%%%%% SUBSECTION %%%%%%%%%%%%%%%%%%%%%%%%%%%%%%%%%%%
%
\subsection{General set of gauge-invariant equations}

A prototypical classical evolution equation
for a real scalar field, $\varphi$, 
interacting with a thermal plasma, is offered by the Langevin equation, 
\be
 {\varphi^{;\mu}_{ }}^{ }_{;\mu} - \Upsilon \, u^{\mu}_{ } \varphi^{ }_{,\mu}
  - V^{ }_{\der\varphi}  
  + \varrho 
 \; = \; 0
 \;. \la{langevin} 
\ee
Here $(...)^{ }_{;\mu}$ is a covariant derivative; 
$(...)^{ }_{,\mu}$ is a normal partial derivative; 
$u^\mu_{ }$ is the plasma four-velocity; 
$\Upsilon$ is a friction coefficient, transporting energy
from~$\varphi$ to the plasma; 
$V$ is the scalar field self-interaction potential, 
with $V^{ }_{\der\varphi} \equiv \partial^{ }_\varphi V$;
$\varrho$ is a stochastic noise term, representing the
energy that $\varphi$ gains from plasma scatterings; 
and we assume the metric signature ($-$$+$$+$$+$). 

Equation~\nr{langevin} should be viewed as a supplement to the 
conservation equations ${T^{ }_{\mu\nu}}^{;\mu}_{ }  = 0$ that dictate
the dynamics of the plasma. Similarly to the ``constitutive relations''
by which we express $T^{ }_{\mu\nu}$ in terms of thermodynamic potentials
and the flow velocity, \eq\nr{langevin} represents the leading terms
of a derivative expansion, describing 
the low-energy dynamics of long-range degrees of freedom
of~$\varphi$. From 
a quantum-mechanical point of view, a non-trivial 
feature is that time reversal is not a symmetry. 
For $\Upsilon > 0$, dissipation and fluctuations drive the 
low-energy modes of $\varphi$ towards the equilibrium state.

The properties of the noise, $\varrho$, are dictated by its autocorrelation
function. The determination of the autocorrelation function is a key part 
of our investigation, and constitutes the contents 
of \se\ref{se:noise}. 

We linearize \eq\nr{langevin} around a background solution, 
$\varphi \equiv \bar\varphi + \delta\varphi$.
We likewise linearize the Einstein equations, 
$\delta G^{ }_{\mu\nu} = 8 \pi G \delta T^{ }_{\mu\nu}$, 
and energy-momentum conservation, 
$\delta {T^{ }_{\mu\nu}}^{;\mu}_{ }  = 0$, 
where $G \equiv 1/\mpl^2$ is Newton's constant, and 
$T^{ }_{\mu\nu}$ is the energy-momentum tensor
describing the inflaton field~$\varphi$ 
and an ideal fluid~\cite[eq.~(2.5)]{alica}.
We account for scalar 
perturbations in the scalar field ($\delta \varphi$), 
temperature ($\delta T$), flow velocity ($\delta v$), 
and metric,  
\be
 g^{ }_{\mu\nu} 
 \; \equiv \;
 a^2_{ }
 \left(
  \begin{array}{cc} 
   -1 - 2 h^{ }_0 & \partial^{ }_i h \\ 
   \partial^{ }_i h & 
   ( 1 - 2 h^{ }_\rmii{D}) \delta^{ }_{ij} + 
   2 \bigl( 
     \partial^{ }_i \partial^{ }_j - \delta^{ }_{ij} \frac{\nabla^2}{3}
     \bigr) \vartheta
  \end{array} 
 \right) %% + \rmO(\delta^2) 
 \;. \la{metric}
\ee
In \eq\nr{metric}, four scalar perturbations have been
included ($h^{ }_0$, $h^{ }_\rmii{D}$, $h$, $\vartheta$).
Through a linear transformation, 
the resulting set of evolution equations can be reduced to a 
$3\times 3$ system
(cf.\ \eqs\nr{dot_R_k}--\nr{dot_S_T} below), 
for the gauge-invariant variables 
\ba
 \R^{ }_\varphi
 & \equiv & 
 - \biggl( 
  h^{ }_\rmii{D} + \frac{\nabla^2\vartheta}{3} 
 \biggr)
  - H\,\frac{\delta\varphi}{\dot{\bar\varphi}}
 \la{def_R_varphi}
 \;, \\ 
%%%
 \R^{ }_v
 & \equiv & 
 - \biggl( 
  h^{ }_\rmii{D} + \frac{\nabla^2\vartheta}{3} 
   \biggr)
 + a  H \, (h - \delta v)
 \;, \la{def_R_v} \\ 
%%%
 \R^{ }_\iT
 & \equiv & 
 - \biggl( 
  h^{ }_\rmii{D} + \frac{\nabla^2\vartheta}{3} 
   \biggr)
 - H\, \frac{\delta T}{\dot{T}}
 \;. \la{def_R_T}
\ea
These variables
are referred to as {\em curvature perturbations}. 
The noise $\varrho$ in \eq\nr{langevin}
% constitutes the only inhomogeneous term, 
% and it 
is treated as being of the same order 
as $\delta\varphi$, $\delta v$ and $\delta T$.  

We can represent the perturbations and the noise
as Fourier modes in spatial directions. There is 
however the delicate point that we want to {\em match} our
solution to quantum-mechanical initial conditions. We then need
a formalism which permits to incorporate both quantum-mechanical
and statistical expectation values. A way to do this is to 
represent all linear perturbations
in the language of a {\em mode expansion}, 
\be
 \R^{ }_{\varphi}
 \; \equiv \;
 \int \! \frac{{\rm d}^3\vec{k}}{\sqrt{(2\pi)^{3}_{ }}} \, 
 \Bigl[ \,
    w^{ }_\rmii{\vec{k}}
    \, \R^{ }_{\varphi{}k}(t) 
    \, e^{ i \vec{k}\cdot\vec{x}}_{}   
  + 
    w^{\dagger}_\rmii{\vec{k}}
    \, \R^{*}_{\varphi{}k}(t) 
    \, e^{ - i \vec{k}\cdot\vec{x}}_{}   
 \, \Bigr]
 \;, \la{mode_expansion}
\ee
where the annihilation and creation operators satisfy
$
 [w^{ }_\rmii{\vec{k}},w^{\dagger}_\rmii{\vec{l}}]
 \equiv
 \delta^{(3)}_{ }(\vec{k-l})
$.
In the text-book interaction picture, the mode expansion is 
written for free fields, and the mode functions are
solved from the corresponding classical field equations.
In our situation, $\R^{ }_\varphi$ interacts with the 
plasma degrees of freedom. To include this physics, 
we adopt a description in 
which the plasma interactions promote the mode functions, 
$\R^{ }_{\varphi{}k}$, to stochastic variables. 

We can now define the quantum-mechanical and stochastic
expectation values. For the quantum side, 
the observable of our interest reads 
\be
 \langle 0 | 
 \R^2_\varphi(t,\vec{x})
 | 0 \rangle 
 \; = \; 
 \int_{-\infty}^{+\infty}
 \! {\rm d}\ln k \, 
 \frac{k^3_{ }}{2\pi^2_{ }}
 | \R^{ }_{\varphi{}k}(t)|^2_{ } 
 \;, \la{pre_P_R_k}
\ee
where $|0\rangle$ refers to the distant-past
(Bunch-Davies) vacuum, 
$ w^{ }_\vec{k} | 0 \rangle = 0$,
from which we assume the initial conditions to 
be sampled. The observed {\rm curvature power spectrum} involves
in addition an 
average over the possible stochastic realizations of the mode functions, 
\be
 \bigl\langle\, 
 \P^{ }_{\R_\varphi}(t,k) 
 \,\bigr\rangle
 \; \equiv\; 
 \frac{k^3_{ }}{2\pi^2_{ }}
 \, 
 \bigl\langle\, 
 | \R^{ }_{\varphi{}k}(t)|^2_{ } 
 \,\bigr\rangle
 \;. \la{P_R_k}
\ee

Given that our equations are linear, 
we are free to rescale variables, 
without changing the form of the equations. 
It is convenient to write the power
spectrum of \eq\nr{P_R_k} directly as the absolute value 
squared of one of our dynamical variables, 
\be
 \bigl\langle\, 
 \P^{ }_{\R^{ }_\varphi}(t,k) 
 \,\bigr\rangle
 \; 
 \underset{\rmii{\nr{def_R_k}}}{
 \overset{\rmii{\nr{P_R_k}}}{=}}
 \; 
 \bigl\langle\,
 |\R^{ }_k(t)|^2_{ }
 \,\bigr\rangle
 \;. \la{P_R}
\ee
To this aim, we define
\ba
 \R^{ }_k 
 & 
 \underset{\rmii{\nr{P_R}}}{
 \overset{\rmii{\nr{P_R_k}}}{\equiv}} 
 &
 \frac{k^{3/2}_{ }}{\sqrt{2\pi^2_{ }}}\,
 \R^{ }_{\varphi{}k} 
 \;, \la{def_R_k} \\[2mm]
%%%%%%
 \E^{ }_v & \equiv & 
 \frac{k^{3/2}_{ }}{\sqrt{2\pi^2_{ }}}\,
 (\bar{e} + \bar{p}) 
 (\R^{ }_{v{}k} - \R^{ }_{\varphi{}k})
 \;, \la{def_S_v} \\[2mm]
%%%%%%
 \E^{ }_\iT & \equiv &
 \frac{k^{3/2}_{ }}{\sqrt{2\pi^2_{ }}}\,
 \bar{e}^{ }_{\der\iT}\, \dot{T}\,
 (\R^{ }_{\iT{}k} - \R^{ }_{\varphi{}k})
 \;, \la{def_S_T}
\ea
where 
$f^{ }_{,\hspace*{0.3mm}x} \equiv \partial^{ }_x f$,  
and dots denote physical time derivatives
($\bar e$ and $\bar p$ are defined below \eq\nr{cal_F}).
The variables $\E^{ }_v$ and $\E^{ }_\iT$ are called 
{\em isocurvature perturbations}, because the first (``curvature'')
term from \eqs\nr{def_R_varphi}--\nr{def_R_T} drops out in the
differences. The equations governing the evolutions of
these perturbations can be found in 
\eqs(3.42)--(3.44) of ref.~\cite{alica},
\ba
 \ddot{\R}^{ }_k
 & = & 
 - \frac{\varrho^{ }_k H}{\dot{\bar\varphi}}
 - \dot{\R}^{ }_k 
    \, \bigl[\, \Upsilon + 2 \F + 3 H \,\bigr]
 -  \R^{ }_k 
    \, \biggl[\, \frac{k^2_{ }}{a^2_{ }} \,\biggr]
 \nn[2mm]
%%%%%%%
 & & \; + \, 
   \E^{ }_v\, 
   \biggl[\,
       \frac{4\pi G ( \Upsilon  +  2 \F )}{H}
   \,\biggr] 
 - \E^{ }_\iT \, 
   \biggl[\,
   \frac{4\pi G}{H}
 \biggl( 1 - \frac{\bar{p}^{ }_{\der\iT}}{\bar{e}^{ }_{\der\iT}}\biggr)
   + \frac{V^{ }_{\der\varphi\iT} 
   + \Upsilon^{ }_{\der\iT} \dot{\bar\varphi} }
     {\dot{\bar\varphi} \, \bar{e}^{ }_{\der\iT}}
   \,\biggr]
 \;, \hspace*{5mm} \la{dot_R_k}
 \\[3mm]
%%%%%%%%
 \dot{\E}^{ }_v
 & = & 
  - \dot{\R}^{ }_k 
    \, \bigl[ \bar{e} + \bar{p} \bigr]
  - \E^{ }_v 
 \biggl[\, 
 3 H + 
 \frac{4 \pi G \dot{\bar\varphi}^2_{ }}{H}
 \,\biggr]
 + 
 \E^{ }_\iT
 \,
 \biggl[\, 
   \frac{\bar{p}^{ }_{\der\iT}}{\bar{e}^{ }_{\der\iT}}
 \,\biggr]
 \;, \la{dot_S_v}
 \\[3mm]
%%%%%%%%
 \dot{\E}^{ }_\iT
 & = & 
 \varrho^{ }_k \, \dot{\bar\varphi}  H
 + \dot{\R}^{ }_k\,
  \biggl[\,
   \Upsilon\dot{\bar\varphi}^2_{ }
  + \frac{8\pi G \bar{e}(\bar{e}+\bar{p})}{H}
  \,\biggr]
 - \R^{ }_k \, 
   \biggl[\, 
     (\bar{e} + \bar{p}) \frac{k^2_{ }}{a^2_{ }}
   \,\biggr]
 \nn[2mm]
%%%%
 &  & \; -\,
 \E^{ }_v \,
 \biggl[\,
  \frac{k^2_{ }}{a^2_{ }} 
 + \frac{4\pi G }{H}
   \biggl(\, \Upsilon \dot{\bar\varphi}^2_{ } 
       + \frac{8 \pi G \bar{e}(\bar{e} + \bar{p})}{H}
   \,\biggr)
 \,\biggr]
 \nn[2mm]
%%%%
 &  & \; + \, 
 \E^{ }_\iT \, 
 \biggl[\,
   \frac{\dot{H} - 4\pi G(\bar{e} + \bar{p})}{H}
  - 3 H \,\biggl( 1 
  + \frac{\bar{p}^{ }_{\der\iT}}{\bar{e}^{ }_{\der\iT}}\,\biggr)
   + \frac{( V^{ }_{\der\varphi\iT} 
   + \Upsilon^{ }_{\der\iT} \dot{\bar\varphi})\,\dot{\bar\varphi}  }
     {\bar{e}^{ }_{\der\iT}}
 \,\biggr]
 \;, \la{dot_S_T}
\ea
where the coefficient $\F$ is defined as
\ba
 \F & \equiv & 
 \frac{H}{\dot{\bar\varphi}}
 \biggl( \frac{\dot{\bar\varphi}}{H} \biggr)^{\textstyle .}_{ }   
 \; = \; 
 \frac{\ddot{\bar\varphi}}{\dot{\bar\varphi}} 
 - \frac{\dot H}{H}  
 \;, \la{cal_F} 
\ea
and $\bar e \equiv e^{ }_r + V$ and $\bar p \equiv p^{ }_r - V$ 
include only the non-derivative parts of the background energy density and
pressure (the derivatives $\dot{\bar\varphi}^2_{ }$ 
are displayed separately). 
By $e^{ }_r$ and $p^{ }_r$ we denote the energy density and pressure
of the plasma (``radiation''). 

For the background evolution, 
we assume that thermal corrections to the 
scalar field potential are negligible, 
$V^{ }_{\der\iT}  \approx 0$.\footnote{%
 If this is not the case, warm inflation model building
 becomes challenging~\cite{warm0}, however the background evolution equations
 themselves are not much more complicated
 (cf.,\ e.g.,\ eqs.~(4.23)--(4.25) of ref.~\cite{alica}). 
 } 
Then the evolution equations read
\ba
  \ddot{ \bar{\varphi} }
 + \bigl( \Upsilon + 3H \bigr)\,\dot{ \bar{\varphi} }
 + V^{ }_{\der\varphi}( \bar{\varphi} )
 & = & 0 
 \;, \la{bg_varphi}
 \\[2mm]
%%%%%
 \dot{e}^{ }_r + 3 H \bigl( e^{ }_r + p^{ }_r % - T  V^{ }_{\der\T}
  \bigr)
 % - T \dot{V}^{ }_{\der\T}  
 & = &
 \Upsilon^{ }
 \dot{\bar\varphi}^2
 \;, \la{bg_T}
 \\[2mm]
%%%
 \sqrt{\frac{8\pi}{3}} 
 \frac{\sqrt{ 
 \tfr{1}{2}\! \dot{\bar\varphi}^2 
  +
  e^{ }_r + V % -T  V^{ }_{\der\T}  
  }}
      { \mpl^{ } }
 & = & 
 H 
 \;. \la{bg_H}
\ea
At late times, 
$\varphi$ starts oscillating if the potential
has positive curvature, and then 
the dynamics can be smoothly matched onto that of an averaged energy
density, $e^{ }_\varphi$; and finally, once $e^{ }_\varphi \ll e^{ }_r$, 
we can drop~$e^{ }_\varphi$ from the evolution equations. These 
transitions are important for quantitatively matching our solution
to a reheated Standard Model plasma (cf.\ \eq\nr{t_1}).  

Through \eqs\nr{dot_R_k}--\nr{dot_S_T}, 
the problem has taken a purely classical appearance. Quantum mechanics
lies in the initial conditions of the mode functions, 
and we return to them in \se\ref{ss:initial}. 
Of course, it should be stressed 
that the corresponding solution does not represent an {\em exact} 
answer to the full quantum-statistical problem. Rather, it is 
an approximate answer, which has the correct quantum-mechanical 
limit when 
the plasma-dependent quantities have not yet played a role, as well as 
the correct classical limit when the system has
equilibrated and all memory 
about initial conditions has been lost. At intermediate
times, the solution smoothly interpolates between these
limits. By tuning the noise, as will be described in \se\ref{se:noise}, 
we can however ensure that the solution is correct also  
deep inside the Hubble horizon, when 
both quantum and statistical fluctuations are present. Given that the 
solution exits the Hubble horizon exponentially fast, 
and freezes out almost immediately after having done so, 
we expect the third regime to guarantee that 
our solution represents a phenomenologically reasonable 
approximation to the full power spectrum.\footnote{%
 It would be interesting to employ the recently popular framework
 of open effective field theories to possibly consolidate this picture
 (cf., e.g., ref.~\cite{oeft} and references therein).
 }

%%%%%%%%%%%%%%%%%%%%%%%%%%% SUBSECTION %%%%%%%%%%%%%%%%%%%%%%%%%%%%%%%%%%%
%
\subsection{Scale hierarchies and simplifications for the weak regime}
\la{ss:scales}

The coefficients that appear in \eqs\nr{dot_R_k}--\nr{dot_S_T}
vary rapidly with time, as dictated by the background solution. 
% and some of them are otherwise intransparent as well. 
It may not
be obvious at first sight which of them are the most important ones. 
However, the system does develop scale hierarchies during its 
evolution, implying that some terms matter more than others. 
Here we recall some of the key notions. 

\bi

\item[(i)]
It is important to know the relations between
the Hubble rate, $H$, the friction coefficient, $\Upsilon$, 
and the temperature, $T$. The first two are themselves functions of  
$\bar\varphi$ and~$T$. The relations between the three are
time-dependent, and model-dependent.  
If $T, \Upsilon \gg H$, we say that we find ourselves in the {\em strong
regime} (of warm inflation). Then the plasma influences both the 
background evolution, and first-order perturbations. If, instead, 
$\Upsilon \ll H \ll T$, we talk about a {\em weak regime}. 
In this case, thermal fluctuations may be important 
(cf.\ the discussion below \eq\nr{res_Omega_ir}), 
but $\Upsilon$ does not influence much the background evolution. 
Finally, if $T,\Upsilon  \ll H$, a thermal plasma could still be present
(provided that its internal equilibration rate satisfies $\Gamma \gg H$, 
which requires strong self-interactions), 
but it does not influence inflationary predictions. 

\item[(ii)]
The coefficient that evolves most rapidly is 
the physical momentum, $k/a$.
At early times, $k/a \gg H$, and the modes are inside
the Hubble horizon. At intermediate times, $k/a \ll H$, and the modes
are outside of the Hubble horizon. Then the modes 
are ``frozen'', i.e.\ $\dot{\R}^{ }_k \approx 0$. At late
times, the modes re-enter inside the Hubble horizon, 
$k/a \gg H$, and the system starts to undergo acoustic oscillations.  

\item[(iii)]
In the course of time, the degree of freedom that carries
the background energy density changes. In the standard picture, 
$V$ is the dominant component at early times; 
$\dot{\bar\varphi}^2_{ }$ and $V$ are of similar magnitudes when 
inflation ends, 
and may remain the most important component for a while; 
ultimately, the plasma energy density takes over, 
and radiation-dominated expansion begins. We may expect 
that perturbations reflect this change. Notably, in the weak regime, even 
if $\E^{ }_v$ and $\E^{ }_\iT$ show a non-trivial evolution, they
do not much affect the dynamics of $\R^{ }_k$; 
and {\it vice versa}, at late times, after Hubble horizon 
re-entry, $\R^{ }_k$ evolves, 
but this dynamics decouples from those of 
$\R^{ }_{v{}k}$ and $\R^{ }_{\iT{}k}$~\cite{alica}.

\ei 

Making use of these scale hierarchies
and restricting ourselves to a specific epoch, we can sometimes
simplify the evolution equations. Let $t^{ }_\rmi{out}$ denote 
a time at which a given mode is well outside of the Hubble horizon. 
Until this moment, 
and if $\varphi$ dominates both the potential and kinetic energy, 
we expect the dynamics to be governed by 
the first line of \eq\nr{dot_R_k}, {\it viz.}
\be
 \ddot{\R}^{ }_k
 +  
    \bigl(\, \Upsilon + 2 \F + 3 H \,\bigr)
    \, 
    \dot{\R}^{ }_k 
 +  
    %% \biggl[\,
      \frac{k^2_{ }}{a^2_{ }} 
    %% \,\biggr]
    \, 
    \R^{ }_k 
 \; 
 \underset{~e^{ }_r \; \ll \; \dot{\bar\varphi}^2_{ } }{
 \overset{~~~t\; \le \; t^{ }_\rmii{out} \vphantom{\big | }}{\approx}} 
 \;\; 
 - \frac{\varrho^{ }_k H}{\dot{\bar\varphi}}
 \;. \la{dot_R_k_simpl}
\ee
This truncation is gauge invariant, 
with the non-trivial information about going over to gauge-invariant
variables contained in the coefficient $\F$ from \eq\nr{cal_F}. 
In terms of the nomenclature from point~(i), we expect
\eq\nr{dot_R_k_simpl} to be valid in the weak regime. 
It gets simplified further
if we invoke one of the hierarchies mentioned under~(ii). 
In \se\ref{se:benchmarks}, we will  
compare the solution obtained from \eq\nr{dot_R_k_simpl} with
the full solution resulting from \eqs\nr{dot_R_k}--\nr{dot_S_T}, 
confirming that the difference becomes large 
in the strong regime 
(cf.\ \fig\ref{fig:histograms} on p.~\pageref{fig:histograms}).

%%%%%%%%%%%%%%%%%%%%%%%%%%% SUBSECTION %%%%%%%%%%%%%%%%%%%%%%%%%%%%%%%%%%%
%
\subsection{Classical representation of the quantum-mechanical initial state}
\la{ss:initial}

With the help of \eqs\nr{def_R_varphi}, \nr{mode_expansion}, 
\nr{def_R_k}, and \nr{dot_R_k_simpl}, 
we can deduce the initial conditions for 
the mode function~$\R^{ }_k$.
Let us denote by $t^{ }_1 \ll t^{ }_\rmi{out}$ 
an early time at which the mode considered is 
well inside the Hubble horizon but non-thermal, with 
$
 k/a^{ }_1 \gg  
 T^{ }_1  
 \,,\,
 \Upsilon^{ }_1 + 2 \F^{ }_1 + 3 H^{ }_1
$, 
where we denote $f^{ }_1 \equiv f(t^{ }_1)$. 
Anticipating that the noise becomes exponentially suppressed
in this regime, 
cf.\ \eq\nr{res_Omega}, \eq\nr{dot_R_k_simpl} is  
dominated by the two terms from its homogeneous part, 
rendering it the harmonic oscillator equation of motion, 
\be
 \ddot{\R}^{ }_k
 +  
    %% \biggl[\,
      \frac{k^2_{ }}{a^2_{ }} 
    %% \,\biggr]
    \, 
    \R^{ }_k 
 \; 
 \overset{~t\; \le \; t^{ }_1 \vphantom{\big | } }{\approx} 
 \;\; 
 0
 \;. \la{dot_R_k_ho}
\ee
This has known trigonometric solutions. Choosing the 
{\em forward-propagating} one, as is appropriate for the 
mode expansion in \eq\nr{mode_expansion}, we can express
the first initial condition as 
\be
 \dot{\R}^{ }_k(t^{ }_1)  
 \; 
 \approx 
 \;
 - i \frac{k}{a^{ }_1}
 \R^{ }_k(t^{ }_1)
 \;. \la{initial_dot_R_k}
\ee
Let us stress that \eq\nr{initial_dot_R_k} requires that 
$\R^{ }_k \in \mathbbm{C}$.

The second initial condition originates from the fact that 
deep inside the Hubble horizon, 
$\delta\varphi$ in \eq\nr{def_R_varphi} should be a local 
canonically normalized field, satisfying 
\be
 [\delta\varphi(t^{ }_1,\vec x)\,,\,\delta\dot{\varphi}(t^{ }_1,\vec y)]
 \;=\;
 \frac{i\,  \delta^{(3)}_{ }(\vec{x-y})}{a_1^3}
 \;. \la{varphi_canonical}
\ee
Here $\vec{x}$ and $\vec{y}$
are comoving rather than local Minkowskian coordinates, 
which produces the division by $a_1^3$.  
From \eq\nr{mode_expansion}, 
making use of the fact that the mode functions are symmetric in 
$\vec k \to -\vec k$, and suppressing the time argument $t^{ }_1$, 
we find 
\ba
 [\R^{ }_\varphi(\vec x)\,,\,\dot{\R}^{ }_\varphi(\vec y)]
 & 
 \underset{\vec k \to -\vec k}{
 \overset{\rmii{\nr{mode_expansion}}}{=}} 
 &  
 \int \! \frac{{\rm d}^3_{ }\vec{k}}{(2\pi)^3_{ }}
 \Bigl\{ 
  \R^{ }_{\varphi{}k} 
  \hspace*{-4mm}
  \overbrace{
  \dot{\R}^{*}_{\varphi{}k}
  }^{\rmii{\nr{initial_dot_R_k}}:\; i \frac{k}{a_1} \R^*_{\varphi{}k} }
  \hspace*{-4mm}
 - 
  \R^{*}_{\varphi{}k} 
  \hspace*{-2mm}
  \overbrace{
  \dot{\R}^{ }_{\varphi{}k}
  }^{ -i \frac{k}{a_1} \R^{ }_{\varphi{}k} }
  \hspace*{-2mm}
 \Bigr\} \, e^{i\vec{k}\cdot(\vec x - \vec y)}_{ }
 \nn[2mm]
%%%%%%
 & 
 \overset{\rmii{\nr{initial_dot_R_k}}}{\approx} 
 & 
 \int \! \frac{{\rm d}^3_{ }\vec{k}}{(2\pi)^3_{ }}
 \, \frac{2 i k}{a^{ }_1} 
 \, |\R^{ }_{\varphi{}k}|^2_{ }
 \, e^{i\vec{k}\cdot(\vec x - \vec y)}_{ }
% \nn[2mm]
%%%%%%
% &
 \; 
  \underset{\rmii{\nr{varphi_canonical}}}{
  \overset{\rmii{\nr{def_R_varphi}}}{\approx}} 
  \; 
 \frac{H_1^2}{\dot{\bar\varphi}_1^2}
 \frac{i\, \delta^{(3)}_{ }(\vec{x-y})}{a_1^3}
 \nn[2mm]
%%%%%
 & \Rightarrow & 
 |\R^{ }_{\varphi{}k}|^2_{ }
 \; \approx \; 
 \frac{H_1^2}{\dot{\bar\varphi}_1^2}
 \frac{1}{2 k a_1^2} 
 \;. \la{R_commutator}
\ea
Including the rescaling by 
$
 k^{3/2}_{ }/\sqrt{2\pi^2_{ } \vphantom{\big | } }
$ 
from \eq\nr{def_R_k}, and choosing a sign, this yields
\be
%%%
 \R^{ }_k(t^{ }_1)
 \;
 \underset{\rmii{\nr{def_R_k}}}{
 \overset{\rmii{\nr{R_commutator}}}{\approx}}
 \;
 \frac{H^{ }_1}{\dot{\bar\varphi}^{ }_1} 
 \frac{ k }{2\pi a^{ }_1} 
 \;. \la{initial_R_k} % \\[2mm] 
\ee
Together with \eq\nr{initial_dot_R_k}, the initial conditions for
$\R^{ }_k$ have thereby been fixed. As we view the initial state
as effectively non-thermal, with $k/a^{ }_1 \gg T^{ }_1$, we also set
$\E^{ }_v(t^{ }_1) \approx \E^{ }_\iT(t^{ }_1) \approx 0$.

%%%%%%%%%%%%%%%%%%%%%%%%%%% SECTION %%%%%%%%%%%%%%%%%%%%%%%%%%%%%%%%%%%%%%
%
\section{What an effective-theory perspective tells about the thermal noise}
\la{se:noise}

The next task is to determine the autocorrelator of the noise 
$\varrho^{ }_k$, appearing in \eqs\nr{dot_R_k}, \nr{dot_S_T} and  
\nr{dot_R_k_simpl}.
We remark that the coefficient $\Upsilon$ can be determined 
independently of the noise, 
from linear response theory~\cite{warm} or more general 
considerations~\cite{db}, and we therefore assume that
its value is known. 
Like in the classic Langevin equation, 
the noise is assumed to be white in time, 
\be
 \bigl\langle\, \varrho^{ }_k(t) \, \varrho^{ }_k(t') \,\bigr\rangle
 \; = \; 
 \Omega^{ }_k \, \delta\bigl(\, t-t' \,\bigr)
 \;. \la{def_Omega_k}
\ee
In the practical implementation we also treat it as Gaussian, 
even if this has formally no influence at the linear order 
to which we restrict ourselves (cf.\ \eq\nr{matrix_formalism}). 

We would like to fix the autocorrelator, $\Omega^{ }_k$, so that
the resulting power spectrum agrees with that predicted
by quantum statistical physics in a certain limit. In practice, 
we consider times early enough so that 
$e^{ }_r \ll \dot{\bar\varphi}^2_{ } $, 
allowing us to make use of the simplified \eq\nr{dot_R_k_simpl}.
We then proceed in four steps:
\bi

\item[(i)]
We choose coordinates and variables such that the problem looks Minkowskian. 
In essence, this means that we go to a local Minkowskian frame, but we
implement this without making any approximation beyond
\eq\nr{dot_R_k_simpl}. We stress that the Minkowskian appearance 
concerns the first-order perturbations, 
whereas the background evolution is kept untouched
(even if it were affected by $\Upsilon$). 

\item[(ii)]
In this frame, we determine the power spectrum with methods of {\em quantum}
statistical physics. This means that the matching coefficients
$\Upsilon$ and $\Omega^{ }_k$ of the classical description are omitted
for a moment. The outcome of this thought experiment represents
the correct answer, in the limit where 
the full system ($\varphi$ + plasma) equilibrates. 
In general, 
only the plasma part is in equilibrium, 
with $T$ defined as its local temperature. 

\item[(iii)]
On the other side, we determine the power spectrum from the effective
{\em classical} description, in the presence of $\Upsilon$ and 
$\Omega^{ }_k$. Again we inspect how the results would look like
in the idealized situation in which we could
wait long enough that the full system equilibrates, even if
in practice probably only the plasma does so. 

\item[(iv)]
Equating the results of the thought experiments in steps (ii) and (iii), 
we can extract~$\Omega^{ }_k$ 
(as mentioned above, 
the other matching coefficient
$\Upsilon$ is assumed known). 

\ei

Before going on with the concrete implementation of these steps, 
it is perhaps useful to answer the following question: 
if we are able to determine the full quantum-mechanical answer, 
from point~(ii), why bother about the effective classical description, from 
point~(iii)? The answer is that the classical framework has 
a broader range of applicability than the equilibrium quantum one, 
describing non-equilibrium situations
as well, and indeed the very process of equilibration of $\varphi$.\footnote{%
 Here we refer to the equilibration of the classical modes, with $k/a \ll T$. 
 The equilibration rate is momentum dependent.
 At weak coupling, the 
 quantum modes, with $k/a \ge T$, equilibrate faster
 than the classical ones, because
 their shorter wavelength permits for faster
 information transfer; as an example, for axion-like coupling,  
 $
  \Upsilon ( m \ll T \le k/a )
  \sim \alphas^3 T^3_{ }/f_a^2
 $, 
 where $\alphas^{ }$ is the coupling of the non-Abelian theory~\cite{bg}. 
 } 
This is essential
for practical applications, as nothing guarantees that $\varphi$ 
actually reaches the equilibrium state before the mode
considered exits the Hubble horizon.

%%%%%%%%%%%%%%%%%%%%%%%%%%%%%%%%%%%%%%%%%%%%%%%%%%%%%%%%%%%%%%%%%%%%%%%%%%
%
\paragraph{(i)~choice of coordinates and variables.}

The goal of the first step is to eliminate $\F$
and $H$ from the friction term in \eq\nr{dot_R_k_simpl}, so that only
the thermal coefficient, $\Upsilon$, remains. 
This can be achieved in three steps. First, we write 
$\Q^{ }_k \equiv - (\dot{\bar\varphi}/H) \, \R^{ }_k$, which 
eliminates $\F$ from the friction 
(but generates new terms, without derivatives acting on~$\Q^{ }_k$). 
Second, we write $\field^{ }_k \equiv a \Q^{ }_k$. This eliminates 
$2 H$ from the friction. Finally, we go to conformal time, $\tau$, with 
\be
 {\rm d}t \; = \; a\, {\rm d}\tau
 \;, \quad
 (...)' 
 \; \equiv \; 
 \partial^{ }_\tau (...) 
 \;=\;
  a \, \partial^{ }_t (...)
 \;. \la{t_tau}
\ee
This eliminates the remaining $H$. The end result takes the form
given in \eq(4.8) of ref.~\cite{alica}, 
\ba
   \bigl(\, 
      \partial_\tau^2
    +  \widehat\Upsilon 
    \partial^{ }_\tau 
    + 
      \overbrace{
      k^2_{ }
    + \widehat\Upsilon \widehat\mu
    + \widehat{m}^2_{ } 
      }^{\;\equiv\;\widehat{\epsilon}{}^{\hspace*{0.4mm}2}_k }
   \,\bigr)
   \, \widehat{\mathcal{Q}}^{ }_k
   & 
    \overset{\rmii{\nr{dot_R_k_simpl}}}{  
    \underset{\rmii{\cite[\eq(4.8)]{alica}}}{ \approx } } 
   &
   \widehat\varrho^{ }_k
   \;, \la{dot_Q_k}
   \\[2mm]
%%%%%%%
 \langle\, 
 \widehat\varrho^{ }_k(\tau) \, \widehat\varrho^{ }_k(\tau') 
 \,\rangle
 & 
 \underset{\rmii{\nr{t_tau}}}{
 \overset{\rmii{\nr{def_Omega_k}}}{=}} 
 & 
 \widehat\Omega^{ }_k \, \delta(\tau-\tau')
 \;, \la{def_hatOmega}
\ea
where we have defined ``conformal'' coefficients and functions as 
\ba
 \widehat \Upsilon & \equiv & a\, \Upsilon
 \;, \quad
 \widehat\varrho^{ }_k \; \equiv \; a^3_{ }\varrho^{ }_k
 \;, \quad
 \widehat\Omega^{ }_k \; \equiv \; a^5_{ }\,\Omega^{ }_k
 \;, \la{rel_Ups} \\[2mm] 
%%%%%
 \widehat \mu & \equiv & 
    - \frac{H}{\bar\varphi\hspace*{0.4mm}'}
      \biggl( \frac{\bar\varphi\hspace*{0.4mm}'}{H} \biggr)'_{ }  
 \; 
 \underset{\rmii{\nr{t_tau},\nr{cal_F}}}{
 \overset{\tau\,\leftrightarrow\,t}{=}} 
 \; 
 - a\, \bigl(\, \F + H \,\bigr)
 \;, \la{def_hatmu}
 \\[2mm]
%%%%%%
 \widehat{m}^2_{ }
 & \equiv & 
     - \frac{H}{\bar\varphi\hspace*{0.4mm}'}
      \biggl( \frac{\bar\varphi\hspace*{0.4mm}'}{H} \biggr)''_{ } 
 \; 
 \underset{\rmii{\nr{t_tau},\nr{cal_F}}}{
 \overset{\tau\,\leftrightarrow\,t}{=}} 
 \; 
 - a^2_{ }\, 
 \bigl(\, \partial^{ }_t + \F + 2 H \,\bigr)
 \bigl(\, \F + H \,\bigr)
 \;. \la{def_hatmm}
\ea
We have also denoted an energy-like variable as
\ba
 \widehat{\epsilon}{}^{\hspace*{0.6mm}2}_k
 & \equiv & 
 k^2_{ }
 + 
 \widehat{\Upsilon}\widehat{\mu} + 
 \widehat{m}^2_{ } 
 \; 
 \equiv
 \; 
 a^2_{ } \epsilon_k^2
 \;, \la{def_hat_eps_k} \\[2mm]
%%%%%
 \epsilon_k^2 
 & = & 
 \frac{k^2_{ }}{a^2_{ }}
 - 
 \bigl(\, \partial^{ }_t + \Upsilon + \F + 2 H \,\bigr)
 \bigl(\, \F + H \,\bigr)
 \;. \la{def_eps_k}
\ea

It may be wondered why we use the notation ${\epsilon}^2_k$ in 
\eq\nr{def_eps_k}, when it is not obvious if the quantity is positive. 
The reason can be understood by working out the second term of \eq\nr{def_eps_k}
explicitly. By using a background evolution equation, we can write
\ba
 \F 
 &
 \underset{\rmii{\nr{bg_varphi}}}{
 \overset{\rmii{\nr{cal_F}}}{=}} 
 & 
 - \frac{V^{ }_{\der\varphi}}{\dot{\bar\varphi}} 
 - \Upsilon 
% \hspace*{-5mm}
% \overbrace{
 -\, 3 H
 - \frac{\dot H}{H}  
% }^{{\rm vanishes~for}\;a\,\to\,{\rm const}}
 \la{calF_expl} 
 \;. 
%%%%%
%  \quad \Rightarrow \quad
%  \dot{\F} 
%  &
%  \underset{\Upsilon\,\sim\,{\rm const}}{
%  \overset{a\,\to\,{\rm const}}{\to}}
%  &
%  - V^{ }_{\der\varphi\varphi} 
%  + \frac{V^{ }_{\der\varphi}}{\dot{\bar\varphi}^2_{ }}
%    \, \ddot{\bar\varphi}
%  \; 
%  \overset{\rmii{\nr{bg_varphi}}}{=} 
%  \; 
%  - V^{ }_{\der\varphi\varphi}  
%  - \biggl( \frac{V^{ }_{\der\varphi}}{\dot{\bar\varphi}} \biggr)^2_{ }
%  - \frac{V^{ }_{\der\varphi}}{\dot{\bar\varphi}} \,\Upsilon 
%  \;, \la{cal_F_mink}
\ea
By taking a time derivative and using again \eq\nr{bg_varphi}, 
we obtain $\dot{\F}$. Inserting into \eq\nr{def_eps_k} and combining 
terms, it can be shown that 
\be
 \epsilon_k^2 
 \; 
 \underset{\rmii{\nr{bg_varphi},\nr{calF_expl}}}{
 \overset{\rmii{\nr{def_eps_k}}}{=}} 
 \; 
 \frac{k^2_{ }}{a^2_{ }}
 + 
 V^{ }_{,\varphi\varphi}
 + 
 \bigl( \Upsilon + 2 \F + 3 H \bigr) \, \frac{\dot{H}}{H}
 + 
 (\partial^{ }_t - H)(\Upsilon + 2 H)
 +
 \frac{\ddot{H}}{H}
 \;. 
 \la{eek}
\ee
{}From here we see that if we keep $a$ and $\Upsilon$ constant, we get 
\ba
 \epsilon_k^2  
 &
 \overset{\rmii{\nr{eek}}\vphantom{\big | }}{
 \underset{a\,,\,\Upsilon\to\,{\rm const}}{\longrightarrow}}
 &
 \frac{k^2_{ }}{a^2_{ }}
 \; + \; 
 V^{ }_{\der\varphi\varphi}
 \;. \la{Vphiphi}
\ea
This is the normal Minkowskian
definition of the energy squared,
with the mass squared given by the curvature of a potential. 
On the other hand, during the slow-roll period, when $\F \ll H$ and 
$\dot{H} \ll H^2_{ }$, and if $V^{ }_{\der\varphi\varphi} < 0$ and $\dot{\Upsilon} < 0$, 
the would-be energy squared reads
\be
 \epsilon^2_k
 \quad
 \underset{\F \,\ll\, H 
          \,,\,  \dot{H}\, \ll\, H^2_{ } 
          \,,\,  V^{ }_{\der\varphi\varphi} \, < \, 0 
          \,,\, \dot{\Upsilon} \, < \, 0  \vphantom{\big | } 
          }{
 \overset{\rmii{\nr{eek}}
          }{<} }
 \quad
 \frac{k^2_{ }}{a^2_{ }}
 - H \, (\Upsilon + 2 H) 
 \;. \la{slow-roll}
\ee
This can become negative as $k/a$ decreases.
The definition of an ``instantaneous'' quantum-mechanical
Hilbert space and statistical physics, 
required for our matching computation, is guaranteed to be sensible only when 
${\epsilon}\hspace*{0.3mm}{}^2_k$ is positive.
However, the results have an analytic continuation
also outside of this domain (see below) 
and, in any case, $\R^{ }_k$ is observed 
to freeze out once a mode exits the Hubble horizon
and $\epsilon^2_k < 0$. 

%%%%%%%%%%%%%%%%%%%%%%%%%%%%%%%%%%%%%%%%%%%%%%%%%%%%%%%%%%%%%%%%%%%%%%%%%%
%
\paragraph{(ii)~quantum-statistical equilibrium power spectrum.}

We now consider the system from \eq\nr{dot_Q_k}. After omitting
the classical matching coefficients from the equations governing
the perturbations (the background evolution is kept fixed), 
the mode equations satisfy
\be
 \field_k^{\hspace*{0.3mm}\prime\prime}
 + 
 \widehat{\epsilon}{}^{\hspace*{0.6mm}2}_k\, 
 \field^{ }_k 
 \;
 \underset{\widehat\Upsilon\,,\,\widehat\Omega^{ }_k\,\to\,0}{
 \overset{\rmii{\nr{dot_Q_k}}}{\approx}} 
 \; 
 0
 \;. 
\ee
The corresponding field operator is like in \eq\nr{mode_expansion}, 
after recalling the rescaling from \eq\nr{def_R_k}, 
\be
 \field^{ }_{\varphi}
 \; \equiv \;
 \int \! \frac{{\rm d}^3\vec{k}}{\sqrt{(2\pi)^{3}_{ }}} \, 
 \Bigl[ \,
    w^{ }_\rmii{\vec{k}}
    \, \field^{ }_{\varphi{}k}(t) 
    \, e^{ i \vec{k}\cdot\vec{x}}_{}   
  + 
    w^{\dagger}_\rmii{\vec{k}}
    \, \field^{*}_{\varphi{}k}(t) 
    \, e^{ - i \vec{k}\cdot\vec{x}}_{}   
 \, \Bigr]
 \;, \quad
 \field^{ }_k 
 \; \equiv \;
 \frac{k^{3/2}_{ }}{\sqrt{2\pi^2_{ }}}\,
 \field^{ }_{\varphi{}k} 
 \;. \la{mode_expansion_Q}
\ee
Forward-propagating mode functions are defined like in 
\eq\nr{initial_dot_R_k}, 
$
 \field^{\hspace*{0.3mm}\prime}_k = 
 - i\, \widehat{\epsilon}^{ }_k\, \field^{ }_k
$, 
and a canonical commutator like 
in \eq\nr{varphi_canonical}, 
$
 [\field^{ }_\varphi(\vec x)
 \,,\,\field^{\hspace*{0.4mm}\prime}_\varphi(\vec y)]
 = 
 i \delta^{(3)}_{ }(\vec x - \vec y)
$.
The result is analogous to \eq\nr{R_commutator}, 
except that we now do {\em not} 
assume $k$ asymptotically large, so that $k$
is replaced by $\widehat{\epsilon}^{ }_k$, leading to 
$
 |\field^{ }_{\varphi{}k}|^2_{ } = 1/(2\, \widehat{\epsilon}^{ }_k)
$.
Subsequently, the vacuum power spectrum is 
\be
 \langle \widetilde 0 | \, \field^{\hspace*{0.3mm}2}_{\varphi}(\vec x) \, 
 | \widetilde 0 \rangle 
 \; = \; 
 \int \! \frac{{\rm d}^3_{ }\vec{k}}{(2\pi)^3_{ }} \, 
 |\field^{ }_{\varphi{}k}|^2_{ }
 \; = \; 
 \int \! \frac{{\rm d}^3_{ }\vec{k}}{(2\pi)^3_{ }} \, 
 \frac{1}{2\, \widehat{\epsilon}^{ }_k}
 \;, \la{P_Q_k_vac}
\ee
where $| \widetilde 0 \rangle$ denotes an instantaneous vacuum
in the local Minkowskian frame. 
These relations are sensible only when $ \widehat{\epsilon}^{ }_k > 0$.

Next, we introduce a finite temperature. The underlying assumption, made
for the purpose of matching, is that $\varphi$ 
has time to equilibrate 
to the instantaneous temperature $T$, defined by the plasma component. 
The equilibration process does not need to be completed in practice, 
because our subsequent classical description also holds
when $\varphi$ is out of equilibrium. 
However, 
the presence of a plasma drives the system {\em towards} equilibrium.  

When we go from vacuum expectation values to quantum statistical physics
within free field theory, 
the mode functions do {\em not} change, but the ensemble is different. 
Considering an excited state with $n^{ }_{\vec q} \ge 0$ quanta of momentum 
$\vec{q}$, non-vanishing matrix elements contributing to a trace are 
\ba
 \langle ... ; n^{ }_{\vec q} ; ... | 
 w^{ }_{\vec k} \, w^\dagger_{\vec q} 
 | ... ; n^{ }_{\vec q}; ... \rangle 
 & = & 
 \delta^{(3)}_{ }(\vec k - \vec q)
 \, 
 \bigl(\,1 + n^{ }_{\vec q} \,\bigr)
 \;, \la{matrix_1}
 \\[2mm]
 \langle ... ; n^{ }_{\vec q} ; ... | 
 w^\dagger_{\vec k} \, w^{ }_{\vec q} 
 | ... ; n^{ }_{\vec q}; ... \rangle 
 & = & 
 \delta^{(3)}_{ }(\vec k - \vec q)
 \, 
  n^{ }_{\vec q} 
 \;. \la{matrix_2}
\ea
The expectation value that includes simultaneously both quantum and
statistical effects is 
\be
 \langle \, \field^{\hspace*{0.4mm}2}_{\varphi}(\vec x) \, \rangle^{ }_\iT
 \; \equiv \; 
 \frac{\tr[ e^{ -\hat H/T}_{ } \,
 \field^{\hspace*{0.4mm}2}_{\varphi}(\vec x) ]}
      {\tr [ e^{ -\hat H/T}_{ } ]}
 \;, 
\ee
where $\hat H$ is the Hamiltonian. Recalling our conformal units, 
the possible occupancies of the momentum 
mode $\vec q$ contribute as 
\ba
 \frac{
 \sum_{n_\vec{q} =0}^{\infty} n^{ }_\vec{q} 
 \exp\bigl( -\frac{ \widehat{\epsilon}^{ }_q n^{ }_\vec{q} }{ a T } \bigr)
 }{
 \sum_{n_\vec{q} =0}^{\infty} 
 \exp\bigl( -\frac{ \widehat{\epsilon}^{ }_q n^{ }_\vec{q} }{ a T } \bigr)
 }
 & = & 
 \frac{(-aT) \frac{{\rm d}}{{\rm d}\widehat{\epsilon}_q}
 \frac{1}{1-\exp\bigl( -\frac{\widehat{\epsilon}_q}{a T} \bigr)} }
 {
 \frac{1}{1-\exp\bigl( -\frac{\widehat{\epsilon}_q}{a T} \bigr)}
 }
 \; = \; 
 \underbrace{
 \frac{1}{\exp\bigl( \frac{\widehat{\epsilon}_q}{a T} \bigr) - 1}
 }_{\,\equiv\, 
 \widehat{n}^{ }_\rmiii{B} \bigl( \frac{\widehat{\epsilon}_q}{a T} \bigr) }
 \;. \la{bose_sum}
\ea
All in all, this yields
\be
 \langle \, \field^{\hspace*{0.4mm}2}_{\varphi}(\vec x) \, \rangle^{ }_\iT 
 \; 
 \overset{\rmii{\nr{mode_expansion_Q}--\nr{bose_sum}}}{=}
 \; 
 \int \! \frac{{\rm d}^3_{ }\vec{k}}{(2\pi)^3_{ }} \, 
 \frac{1}{2\, \widehat{\epsilon}^{ }_k}
 \biggl[\, 1 + 
 2\, \widehat{n}^{ }_\iB \Bigl( \frac{\widehat{\epsilon}_k}{a T} \Bigr)
 \,\biggr]
 \;. \la{P_Q_k_thermal}
\ee
The power spectrum, 
$|\field^{ }_k|^2_{ }$,  
is given by the integrand 
of \eq\nr{P_Q_k_thermal}
after carrying out the (trivial)
angular integrals in spherical coordinates, 
which yields the same volume element 
$k^3_{ }/ (2 \pi^2_{ })$ as seen in \eq\nr{mode_expansion_Q}, 
\be
 |\field^{ }_k|^2_{ }
 \underset{\rmii{\nr{P_Q_k_thermal}}}{
 \overset{\rmii{\nr{mode_expansion_Q}}}{=}}
 \frac{k^3_{ }}{4 \pi^2_{ } \, \widehat{\epsilon}^{ }_k}
 \biggl[\, 1 + 
 2\, \widehat{n}^{ }_\iB \Bigl( \frac{\widehat{\epsilon}_k}{a T} \Bigr)
 \,\biggr]
 \;. \la{P_Q_qm}
\ee

%%%%%%%%%%%%%%%%%%%%%%%%%%%%%%%%%%%%%%%%%%%%%%%%%%%%%%%%%%%%%%%%%%%%%%%%%%
%
\paragraph{(iii)~classical-statistical equilibrium power spectrum.}

Next, we return to \eq\nr{dot_Q_k} and solve it (classically)
in the presence of $\hat\Upsilon$ and $\varrho^{ }_k$. Furthermore, 
we again assume that equilibration to a temperature $T$ 
happens faster than the variation of the parameter values 
($\widehat\Upsilon$, $\widehat{\epsilon}^{ }_k$). 
Under this assumption, 
we determine the classical approximation to $|\field^{ }_k|^2_{ }$.

Equation~\nr{dot_Q_k} has a general homogeneous solution, and a special
solution of the inhomogeneous equation. The homogeneous solution is 
identified with the quantum-mechanical initial conditions, which fix
the two integration constants. To infer the thermal modifications, 
because of the assumed equilibration, 
we can wait long enough that the homogeneous solution has dissipated
away. Then we only need to determine the special solution, and this can 
be done with the help of a retarded Green's function. 
Let us illustrate how this goes. 

Inserting $\delta(\tau - \tau^{ }_1)$
as the right-hand side, 
the solution of \eq\nr{dot_Q_k} is given by 
\be
 G^{ }_k(\tau,\tau^{ }_1)
 \; 
 \equiv
 \; 
 \int \! \frac{{\rm d}\omega}{2\pi} \, 
 e^{ -i\omega(\tau - \tau^{ }_1)}_{ } \, 
 G^{ }_k(\omega)
 \;, \quad
 G^{ }_k(\omega) 
 \; 
 \equiv 
 \; 
 \frac{1}{-\omega^2_{ } + \widehat\Upsilon\, (-i\omega)
 + \widehat{\epsilon}{}^{\hspace*{0.6mm}2}_k}
 \;. \la{G_k}
\ee
The poles lie at 
\be
 \omega 
 \;=\;
  - \frac{i\widehat\Upsilon}{2} \pm X
 \;, \quad
 X \; \equiv \; 
 \sqrt{ \widehat{\epsilon}{}^{\hspace*{0.6mm}2}_k 
  - \frac{\widehat\Upsilon^2_{ }}{4}}
 \;. \la{poles}
\ee
For $\widehat\Upsilon > 0$ and 
$
 \widehat{\epsilon}{}^{\hspace*{0.6mm}2}_k > 0 
$,
so that $|\im X| < \widehat\Upsilon / 2$,
the poles are in the lower half-plane. On the other hand, for 
$\tau - \tau^{ }_1 < 0$, we can close the 
$\omega$-contour in \eq\nr{G_k} in the upper half-plane, 
because the $\omega$-integrand
is exponentially suppressed there. As there are no poles
in the upper half-plane, 
the integral vanishes, i.e.\  
$G^{ }_k(\tau,\tau^{ }_1) = 0$ for $\tau < \tau^{ }_1$. 
Therefore, \eq\nr{G_k} defines a 
{\em retarded} Green's function
(i.e.\ it is non-zero only for $\tau > \tau^{ }_1$).

With the help of $G^{ }_k$ from \eq\nr{G_k}, and assuming 
that $\widehat\Upsilon$ and  
$\widehat \epsilon^{\hspace*{0.5mm}2}_{k}$
are to a good approximation constant 
during the interval considered, a special
solution of \eq\nr{dot_Q_k} is given by
\ba
 \field^{ }_k(\tau) 
 & 
 \underset{\rmii{\nr{G_k}}}{
 \overset{\rmii{\nr{dot_Q_k}}}{=}} 
 & 
 \int_{\tau_1} 
 G^{ }_k(\tau,\tau^{ }_1) \, \widehat{\varrho}^{ }_k(\tau^{ }_1)
 \; 
 = 
 \; 
 \int_{\tau^{ }_1,\omega^{ }_1} \!\!
 e^{-i \omega^{ }_1 (\tau - \tau^{ }_1) }_{ }
 G^{ }_k(\omega^{ }_1) \, \widehat{\varrho}^{ }_k(\tau^{ }_1) 
 \;, \\[2mm]
%%%%%%
 & & 
 \int_{\tau^{ }_1} 
 \;
 \equiv 
 \; 
 \int_{-\infty}^{+\infty} \! {\rm d}\tau^{ }_1
 \;, \quad
 \int_{\omega^{ }_1} 
 \;
 \equiv 
 \; 
 \int_{-\infty}^{+\infty} \! \frac{{\rm d}\omega^{ }_1}{2\pi} \, 
 \;. \la{special_soln}
\ea
Inserting 
\be
 \field^{*}_k(\tau) 
 \; = \; 
 \int_{\tau^{ }_2,\omega^{ }_2} \!\! 
 e^{i \omega^{ }_2 (\tau - \tau^{ }_2) }_{ } 
 G^{\hspace*{0.3mm}*}_k(\omega^{ }_2) \,
 \widehat{\varrho}^{\hspace*{0.5mm}*}_k(\tau^{ }_2) 
 \;,  \la{conjugate}
\ee
the power spectrum becomes 
\ba
 |\field^{ }_k(\tau)|^2_{ }
 \hspace*{-4mm}
 & 
 \underset{\rmii{\nr{conjugate}}}{
 \overset{\rmii{\nr{special_soln}}}{=}}
 & 
 \hspace*{-4mm}
 \int^{ }_{\tau^{ }_1,\,\tau^{ }_2,\,\omega^{ }_1,\,\omega^{ }_2}
 \hspace*{-4mm}
 e^{i\omega^{ }_2(\tau - \tau^{ }_2) - i \omega^{ }_1(\tau - \tau^{ }_1) }_{ }
 \, 
 G^{ }_k(\omega^{ }_1) 
 G^{\hspace*{0.3mm}*}_k(\omega^{ }_2) 
 \; 
 \overbrace{
 \bigl\langle\,
 \widehat{\varrho}^{ }_k(\tau^{ }_1) 
 \widehat{\varrho}^{\hspace*{0.5mm}*}_k(\tau^{ }_2) 
 \,\bigr\rangle
 }^{\rmii{\nr{def_hatOmega}}\,:\;
 \widehat\Omega^{ }_k\delta(\tau^{ }_1 - \tau^{ }_2)}
 \nn[2mm]
%%%%%%%
 \hspace*{-4mm}
 & = & 
 \hspace*{-4mm}
 \widehat{\Omega}^{ }_k
 \int^{ }_{\tau^{ }_1,\,\omega^{ }_1,\,\omega^{ }_2}
 \hspace*{-4mm}
 e^{i(\omega^{ }_2-\omega^{ }_1)(\tau - \tau^{ }_1) }_{ }
 \, 
 G^{ }_k(\omega^{ }_1) 
 G^{\hspace*{0.3mm}*}_k(\omega^{ }_2) 
 \nn[2mm]
%%%%%%%
 \hspace*{-4mm}
 & = & 
 \hspace*{-4mm}
 \widehat{\Omega}^{ }_k
 \int^{ }_{\omega^{ }_1}
 G^{ }_k(\omega^{ }_1) 
 G^{\hspace*{0.3mm}*}_k(\omega^{ }_1) 
 \nn[2mm]
%%%%%%%
 \hspace*{-4mm}
 & 
 \underset{\rmii{\nr{poles}}}{
 \overset{\rmii{\nr{G_k}}}{=}} 
 & 
 \hspace*{-4mm}
 \int_{-\infty}^{+\infty} \! \frac{{\rm d}\omega^{ }_1}{2\pi} \, 
 \frac{ \widehat{\Omega}^{ }_k }{
 \bigl( \omega^{ }_1 + \frac{i \widehat\Upsilon}{2} + X \bigr)
 \bigl( \omega^{ }_1 + \frac{i \widehat\Upsilon}{2} - X \bigr)
 \bigl( \omega^{ }_1 - \frac{i \widehat\Upsilon}{2} + X \bigr)
 \bigl( \omega^{ }_1 - \frac{i \widehat\Upsilon}{2} - X \bigr)
 }
 \nn[2mm]
%%%%%%%
 \hspace*{-4mm}
 & 
 \underset{\rmii{ }}{
 \overset{|{\rm Im} X|\, < \,\widehat\Upsilon / 2\;}{=}} 
 & 
 \hspace*{0mm}
 \frac{\widehat \Omega^{ }_k}{2 X \widehat{\Upsilon}}
 \biggl(\,
  \frac{1}{i \widehat\Upsilon + 2 X}
 + 
  \frac{1}{-i \widehat\Upsilon + 2 X}
 \,\biggr)
 \;
 \overset{\rmii{\nr{poles}}}{=}
 \; 
 \frac{\widehat \Omega^{ }_k}{2 \widehat{\Upsilon}}
 \,
 \frac{1}{\widehat{\epsilon}{}^{\hspace*{0.6mm}2}_k}
 \;. \la{P_Q_cl}
\ea

%%%%%%%%%%%%%%%%%%%%%%%%%%%%%%%%%%%%%%%%%%%%%%%%%%%%%%%%%%%%%%%%%%%%%%%%%%
%
\paragraph{(iv)~matching between quantum and classical descriptions.}

We compare \eqs\nr{P_Q_qm} and \nr{P_Q_cl}, recalling that
the temperature-independent first term of \eq\nr{P_Q_qm} corresponds
to the homogeneous solution, which was excluded from \eq\nr{P_Q_cl}. 
In other words, matching requires
\be
 \underbrace{
 \frac{k^3_{ }}{2\pi^2_{ }\,\widehat{\epsilon}^{ }_k}
 \,\widehat{n}^{ }_\iB \Bigl( \frac{\widehat{\epsilon}_k}{a T} \Bigr)
 }_{{\rm from}\; \rmii{\nr{P_Q_qm}} }
 \; 
 \simeq
 \; 
 \underbrace{
 \frac{\widehat \Omega^{ }_k}{2 \widehat{\Upsilon}}
 \,
 \frac{1}{\widehat{\epsilon}{}^{\hspace*{0.6mm}2}_k }
 }_{{\rm from}\; \rmii{\nr{P_Q_cl}} }
 \quad 
 \Rightarrow 
 \quad
 \widehat \Omega^{ }_k 
 \; \simeq \; 
 2 \widehat\Upsilon\, 
 \widehat{\epsilon}\hspace*{0.4mm}{}^{ }_k
 \,\widehat{n}^{ }_{\iB} 
 \Bigl( \frac{\widehat{\epsilon}_k}{a T} \Bigr)
 \,\frac{k^3_{ }}{2\pi^2_{ }}
 \;. \la{res_hat_Omega}
\ee
Going back to physical units, 
via \eqs\nr{rel_Ups}--\nr{def_eps_k}, then
yields
\be
 \boxed{
 \Omega^{ }_k
 \; 
 \underset{\rmii{\nr{rel_Ups}--\nr{def_eps_k}}}{
 \overset{\rmii{\nr{res_hat_Omega}}}{\simeq}}
 \; 
 2 \Upsilon
 \, \epsilon^{ }_k 
 \, n^{ }_{\iB}
 \bigl( \epsilon^{ }_k \,\bigr)
 \frac{ ( {k}/{a} )^3_{ }}{2\pi^2_{ }}
 }
 \;, \la{res_Omega}
\ee
where the Bose distribution is defined as
$
 n^{ }_{\iB}(x) \equiv 1/( e^{x/T}_{ }-1 )
$, 
and $\epsilon^{ }_k$ is from \eq\nr{def_eps_k}.

Let us elaborate on how \eq\nr{res_Omega} should be interpreted
in different regimes: 

\bi

%%%%%%%%%%%%%%%%%%%%%%%%%%%%%%
\item
$ 
  \displaystyle
  %\boxed{
  \epsilon^{ }_k \ge T \;.
  %}
$
In this domain, the noise autocorrelator from \eq\nr{res_Omega} is 
exponentially suppressed. Then the evolution reflects 
quantum-mechanical initial conditions. 

%%%%%%%%%%%%%%%%%%%%%%%%%%%%%%
\item
$
  \displaystyle
  %\boxed{
  0 < \epsilon^{ }_k  \ll T \;.
  %}
$
Because of the negative terms in \eq\nr{slow-roll}, 
$\epsilon^{ }_k$ decreases as the mode considered is close
to exiting the Hubble horizon. 
In this domain, we can approximate 
\be
 n^{ }_{\iB}(x)
 \;
 \overset{|x|\,\ll\,T}{\approx}
 \; 
 \frac{T}{x} 
 \quad
 \Rightarrow
 \quad
 \Omega^{ }_k 
 \; 
 \underset{\epsilon^{ }_k\,\ll\,T}{
 \overset{\rmii{\nr{res_Omega}}}{\approx}}
 \; 
 2\hspace*{0.3mm} \Upsilon\hspace*{0.3mm} T
 \, \frac{(k/a)^3_{ }}{2\pi^2_{ }}
 \;. \la{res_Omega_ir}
\ee
This corresponds to the textbook fluctuation-dissipation relation 
(modified by our convention of including 
the phase-space volume element in $\Omega^{ }_k$). 
According to \eq\nr{slow-roll},  
\eq\nr{res_Omega_ir} is a good approximation
to \eq\nr{res_Omega} % (and thus not exponentially suppressed) 
in the domain 
$
 k^2_{ }/a^2_{ }\ll (\Upsilon + 2 H) H + T^2_{ } 
$.
Thermal fluctuations have a large effect 
before freeze-out, if 
$
 T^2_{ }\gg (\Upsilon + 2 H) H
$.

%%%%%%%%%%%%%%%%%%%%%%%%
\item
$
  \displaystyle
  %\boxed{
  \epsilon_k^2  \le 0 \;.
  %}
$
When $\epsilon^2_k$ becomes negative, our derivation is not 
physically sensible. That said, 
\eq\nr{res_Omega_ir} is independent of $\epsilon^2_k$, and
can be employed as an extrapolation into this 
domain. Mathematically speaking, it represents a good approximation
to the analytic continuation of \eq\nr{res_Omega} as long as 
$|\epsilon^{ }_k| < 2\pi T$, before $\nB^{ }$ develops 
additional poles along the imaginary axis. Physically speaking, 
the rapidly decreasing $\sim (k/a)^3_{ }$ guarantees that the noise
only plays a minor role outside of
the Hubble horizon. Therefore, we employ 
\eq\nr{res_Omega_ir} in the domain $\epsilon_k^2 < 0$.

\ei

%%%%%%%%%%%%%%%%%%%%%%%%%%% SECTION %%%%%%%%%%%%%%%%%%%%%%%%%%%%%%%%%%%%%%
%
\section{Numerical method for the stochastic evolution}
\la{se:numerics}

The noise autocorrelator in \eq\nr{res_Omega}
is a strongly varying function of time, but it does not 
depend on the dynamical variable that it affects, $\R^{ }_k$. 
Therefore, the stochastic evolution equations can be discretized
by making use of the standard 
It\^{o} or Stratonovich schemes. For us the It\^{o}
scheme turned out to be sufficient, so we describe its practical 
implementation, as well as the simplifications 
that are possible if only
the power spectrum is required
(in contrast to the full probability distribution). 
The same procedure works both for 
the full dynamics from \eqs\nr{dot_R_k}--\nr{dot_S_T}, 
and for the simplified evolution from \eq\nr{dot_R_k_simpl}.  

Given that the coefficients 
vary rapidly with time, it is important to choose a 
dynamically changing time step. 
Let us assemble our variables into a vector, 
\be
 \V \; \equiv \; 
 \bigl(
  \R^{ }_k \; \dot\R^{ }_k \; \E^{ }_v \; \E^{ }_\iT \bigr)^\rmii{T}_{ }
 \;, \la{def_V}
\ee
where $(...)^\rmii{T}_{ }$ denotes a transpose. 
Regularizing the poles appearing 
in \eq\nr{dot_R_k_simpl} as in ref.~\cite{alica},
\be
 \F
 \;
 \overset{\rmii{\nr{cal_F}}}{\to}
 \; 
 \frac{\ddot{\bar\varphi}}{\dot{\bar\varphi} + i \delta } - \frac{\dot H}{H}
 \;, \quad
 - \frac{\varrho^{ }_k H}{\dot{\bar\varphi}}
 \; \to \; 
 - \frac{\varrho^{ }_k H}{\dot{\bar\varphi} + i \delta}
 \;, \la{regularization}
\ee
our continuum equation has the form 
\ba
 \dot{\V} 
 &
 \underset{\rmii{\nr{def_V}}}{
 \overset{\rmii{\nr{dot_R_k_simpl}}}{=}}
 &
 \overbrace{
 \left(
 \begin{array}{ccc} 
  0 & 1 & 0 \quad 0 \\ 
  -\frac{k^2_{ }}{a^2_{ }} &  - (\Upsilon + 2 \F + 3 H ) & \hdots \\ 
 \vdots & \vdots & \ddots
 \end{array} 
 \right)
 }^{ \,\equiv\, M}
 \, \V
 +  
 \N
 \;, \la{dot_V}
 \\[2mm]
%%%%%%%
 \bigl\langle\, \N^{ }_2(t)\, \N^{ }_2(t') \,\bigr\rangle
 & 
 \underset{\rmii{\nr{res_Omega},\nr{regularization}}}{
 \overset{\rmii{\nr{def_Omega_k}}}{=}} 
 & 
 \underbrace{
 2 \Upsilon
 \, \epsilon^{ }_k 
 \, n^{ }_{\iB}
 \bigl( \epsilon^{ }_k \,\bigr)
 \frac{ ( {k}/{a} )^3_{ }}{2\pi^2_{ }}
 \frac{H^2_{ }}{(\dot{\bar\varphi} + i \delta)^2_{ }}
 }_{\,\equiv\, W^{ }_{22} % \;{\rm [GeV^3_{ }]}
   }
 \; \delta(t - t')
 \;. \la{def_W}
\ea
In \eq\nr{dot_V}, we have shown explicitly only the parts originating
from \eq\nr{dot_R_k_simpl}, with the dots denoting the further terms
from \eqs\nr{dot_R_k}--\nr{dot_S_T}.
The matrix $M$ is called the drift term, 
$\N$ the noise term, and $W$ the noise autocorrelator. 

A simple discretization of \eq\nr{dot_V}
amounts to 
\be
 \frac{\V^{ }_{i+1} - \V^{ }_i}{\epsilon^{ }_i} 
 \; = \; 
 M^{ }_i \V^{ }_i + \frac{\N^{ }_i}{\sqrt{\epsilon^{ }_i}}
 \;, \quad
 \bigl\langle\,
   \N^\rmii{\,T}_{i} \N^{ }_{j}
 \,\bigr\rangle
 \; = \; 
 W^{ }_i\, \delta^{ }_{ij}
 \;, \la{discr_V}
\ee
where $f^{ }_i \equiv f(t^{ }_i)$ and
$t^{ }_{i+1} - t^{ }_i \equiv \epsilon^{ }_i$. 
This yields the It\^{o} evolution,
\be
 \V^{ }_{i+1} = \bigl(\, 1 + \epsilon^{ }_i M^{ }_i \,\bigr)
 \, \V^{ }_i + \sqrt{\epsilon^{ }_i} \, \N^{ }_i
 \;. \la{ito_1} 
\ee

A crucial question is how to choose $\epsilon^{ }_i$, so that the 
evolution is stable and close to the correct continuum limit without
an enormous computational cost. We base the criterion on the drift 
term, setting 
\be
 \epsilon^{ }_i \; = \; \frac{\mbox{\tt tol}}
 {\max \{ |\lambda^{ }_{i}|\} }
 \;, \quad \mbox{\tt tol} \ll 1.0
 \;, \la{tolerance} 
\ee
where $\lambda^{ }_i$ are the (complex) eigenvalues of $M^{ }_i$, 
and a sufficient tolerance is   
$\mbox{\tt tol} \lsim 10^{-3}_{ }$.

The matrix $M$ in \eq\nr{dot_V} and the noise 
autocorrelator $W$ in \eq\nr{def_W} are functions of the background
solution.  
Therefore, for a practical implementation, 
we first determine the background solution 
from \eqs\nr{bg_varphi}--\nr{bg_H}. The solution is smooth, and can be 
stored in an array. 
With this information, \eq\nr{ito_1} is straightforward to evolve, 
with the (complex) Gaussian noise generated according to the variance
given in \eq\nr{discr_V}. In other words, if $\sigma^{ }_i$ is a Gaussian
random variable, with variance $\langle \sigma_i^2 \rangle = 1$, 
then\footnote{%
  In practice, we take $t/t^{ }_\rmii{ref}$
  as the integration variable, 
  where $t^{ }_\rmii{ref}$ is defined in \eq\nr{t_ref}.
  Then $M$ and $W$ get rescaled by $t^{ }_\rmii{ref} = H^{-1}_\rmii{ref}$. 
  We note that $\N$
  in \eq\nr{dot_V} gets rescaled
  by $H^{-1}_\rmii{ref}$, so naively it looks like 
  there is $H^{-2}_\rmii{ref}$ in 
  the autocorrelator, however we also have to write 
  $
    H^{-1}_\rmii{ref} \delta(t-t') 
  = \delta (t/t^{ }_\rmii{ref} - t'/t^{ }_\rmii{ref})
  $, 
  whereby only one $H^{-1}_\rmii{ref}$ is left over. 
  In other words, $\N^{ }_{2i}$ in \eq\nr{calN_i} is rescaled
  by $1/\sqrt{H^{ }_\rmii{ref}}$.
%  The same conclusion
%  is obtained if we consider a rescaling of
%  $\epsilon^{ }_i$ in \eq\nr{ito_1}.
  \label{rescaling}
 } 
\be
 \N^{ }_{2i}
 \;
 \underset{\rmii{\nr{discr_V}}}{
 \overset{\rmii{\nr{def_W}}}{=}}
 \;
 -\,
 \sigma^{ }_i 
 \;  
 \biggl[\;
 \sqrt{ 
 2 \Upsilon
 \, \epsilon^{ }_{k} 
 \, n^{ }_{\iB}
 \bigl( \epsilon^{ }_{k} \,\bigr)
 }
 \;
 \frac{ ( {k}/{a} )^{3/2}_{ }}{\sqrt{ 2\pi^2_{ } } }
 \;
 \frac{H^{ }_{ }}{\dot{\bar\varphi} + i \delta}
 \;\biggr]^{ }_{t \; = \; t^{ }_i}
 \;. 
 \la{calN_i}
\ee
The function 
$
 \epsilon^{ }_{k} 
 \, n^{ }_{\iB}
 \bigl( \epsilon^{ }_{k} \,\bigr)
$
is extrapolated to the domain of non-positive arguments
according to \eq\nr{res_Omega_ir}. We remark that, 
when $\bar\varphi$ starts oscillating, the coefficient $\F$
needs to be regularized according to \eq\nr{regularization}, 
and $\epsilon^2_k$ from \eq\nr{eek} obtains an imaginary part.
There is no principal problem with the variables 
becoming complex (the perturbations
have been complexified by the initial 
conditions, cf.\ \eq\nr{initial_dot_R_k}).  
Moreover, in our case the freeze-out dynamics takes place
before $\bar\varphi$ starts oscillating, and the regularization
plays no practical role. 

\vspace*{3mm}

If we are only interested in the power spectrum (cf.\ \eq\nr{P_R}), 
the stochastic evolution can be greatly simplified. Taking the 
conjugate of \eq\nr{ito_1}, multiplying with \eq\nr{ito_1}, and noting
that $\V^{ }_i$ and $\N^{ }_i$ are uncorrelated because 
$\V^{ }_i$ is affected only by $\N^{ }_{i-1}$, we find
\be
 \bigl\langle\, 
 \V^{ }_{i+1} \hspace*{0.3mm} \V^\dagger_{i+1}
 \,\bigr\rangle
 \; 
 \overset{\rmii{\nr{ito_1}}}{=} 
 \; 
 \bigl(\, 1 + \epsilon^{ }_i M^{ }_i \,\bigr)
 \,\bigl\langle\,
 \V^{ }_{i} \hspace*{0.3mm} \V^\dagger_{i}
 \,\bigr\rangle\,
 \bigl(\, 1 + \epsilon^{ }_i M^{\dagger}_i \,\bigr)
 + 
 \epsilon^{ }_i 
 \,\bigl\langle\,
 \N^{ }_{i} \hspace*{0.3mm} \N^\dagger_{i}
 \,\bigr\rangle\, 
 \;. \la{matrix_formalism}
\ee
The initial conditions from \eqs\nr{initial_dot_R_k} and
\nr{initial_R_k} reside in a $2\times 2$ subblock of 
$ 
 \langle \V^{ }_{0} \hspace*{0.3mm} \V^\dagger_{0} \rangle
$, 
and the stochastic noise makes a 
$2\times 2$ subblock matrix out of 
$
 \langle \N^{ }_{i} \hspace*{0.3mm} \N^\dagger_{i} \rangle
$, 
after inserting $\langle \sigma_i^2 \rangle = 1$.
Thereby the evolution of the stochastic expectation
values is deterministic. This approach was referred to as
a {\em matrix formalism} in refs.~\cite{garcia,ballesteros}, 
who first demonstrated its effectiveness 
in the warm inflation context.
Our scan results in 
\figs\ref{fig:fit}--\ref{fig:Tmax}
are based on \eq\nr{matrix_formalism}.

%%%%%%%%%%%%%%%%%%%%%%%%%%% SECTION %%%%%%%%%%%%%%%%%%%%%%%%%%%%%%%%%%%%%%
%
\section{Benchmarks and overall features}
\la{se:benchmarks}

In this section we 
illustrate features 
of the solution of our evolution equations 
from \se\ref{se:numerics}
with a few
representative benchmarks. Subsequently,  
in \ses\ref{se:scans} and \ref{se:sm}, 
we report our first physics applications, 
based on comprehensive scans of such solutions.  
We recall that, when $k\ll aH$,
all curvature perturbations coincide; 
in that domain we streamline the notation by 
dropping the subscript from $\R^{ }_\varphi$ or $\R^{ }_k$, 
so that we refer to the power spectrum as $\P^{ }_\R$. 

For a transparent discussion, it is helpful to 
keep the parameters of the inflaton potential, 
the characteristics of the radiation plasma, 
and the initial conditions,
fixed throughout the scans;  
they are chosen as explained in appendix~\ref{se:V}.
The quantities varied are the friction coefficient, $\Upsilon$, 
and the momentum mode considered, $k$. Let us anticipate
that in phenomenological scans, it is necessary to vary the
parameters of the inflaton potential as well, and indeed we will do 
this for our second application, in \se\ref{se:sm}.

For the friction coefficient, we employ a parametrization
motivated by the structure originating from 
an axion-like coupling between the inflaton field and  
a non-Abelian plasma, with the latter assumed to be 
in kinetic and helicity equilibrium, 
\be
 \Upsilon 
 \; \equiv \; 
 \frac{
    \kappa^{ }_\iT\, (\pi T)^3_{ } + 
    \kappa^{ }_m\, m^3_{ }
 }{(4\pi)^3_{ } f_a^2}
 \;.  \la{Upsilon_ansatz}
\ee
The parameters 
$m$ and $f^{ }_a$ are the same as in \eq\nr{V}.
In a microscopic derivation, $\kappa^{ }_\iT$
and $\kappa^{ }_m$ depend on the coupling $\alphas^{ }$ of the non-Abelian 
theory, and on logarithms of $T$ and $m$, and realistic values are of order  
$
 \kappa^{ }_\iT, \kappa^{ }_m \sim 1 
$~\cite{warm}. 
Moreover, a thorough study reveals
a richer functional form than in \eq\nr{Upsilon_ansatz}~\cite{clgt}.
However, values of order unity 
lead to very small thermal effects for CMB observables
with the potential from \eq\nr{V}~\cite{zell}.
In order to span trajectories 
with large thermal corrections,  
we scan the domains
$\kappa^{ }_\iT \sim 10^{7.0 ... 10.5}_{ }$ and
$\kappa^{ }_m \sim 10^{9.0 ... 17.5}_{ }$.
The upper bounds yield solutions deep in the strong 
regime, which become increasingly costly to handle
numerically, 
because the dynamics is dominated by fast acoustic
oscillations of the radiation plasma 
(cf.\ \fig\ref{fig:cases_curvature} 
on p.~\pageref{fig:cases_curvature}). 

%%%%%%%%%%%%%%%%%%%%%%%%%%%%%%%%% FIGURE %%%%%%%%%%%%%%%%%%%%%%%%%%%%%%%%%
\begin{figure}[t]

 \hspace*{-0.1cm}
 \centerline{%
  \epsfysize=5.0cm\epsfbox{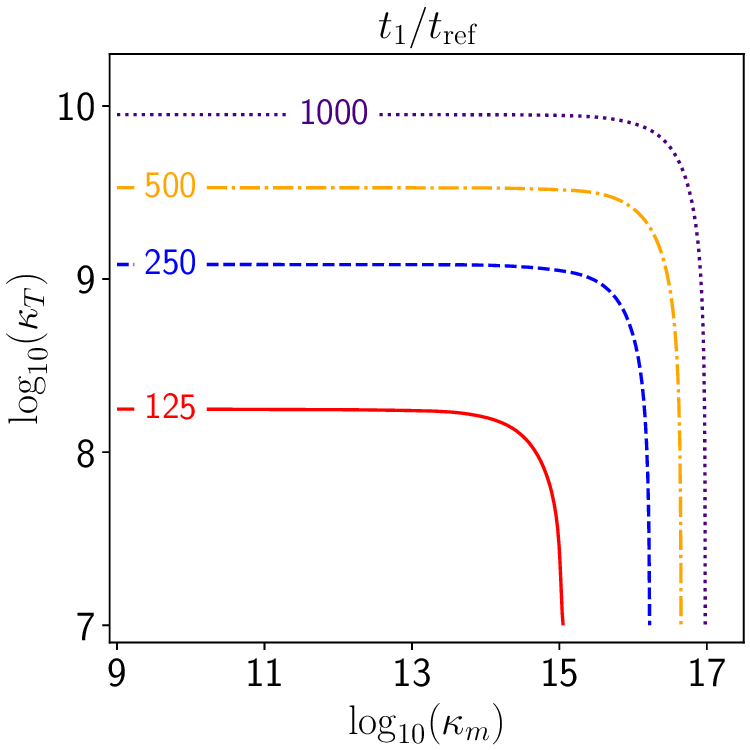}%
  \hspace{0.2cm}%
  \epsfysize=5.0cm\epsfbox{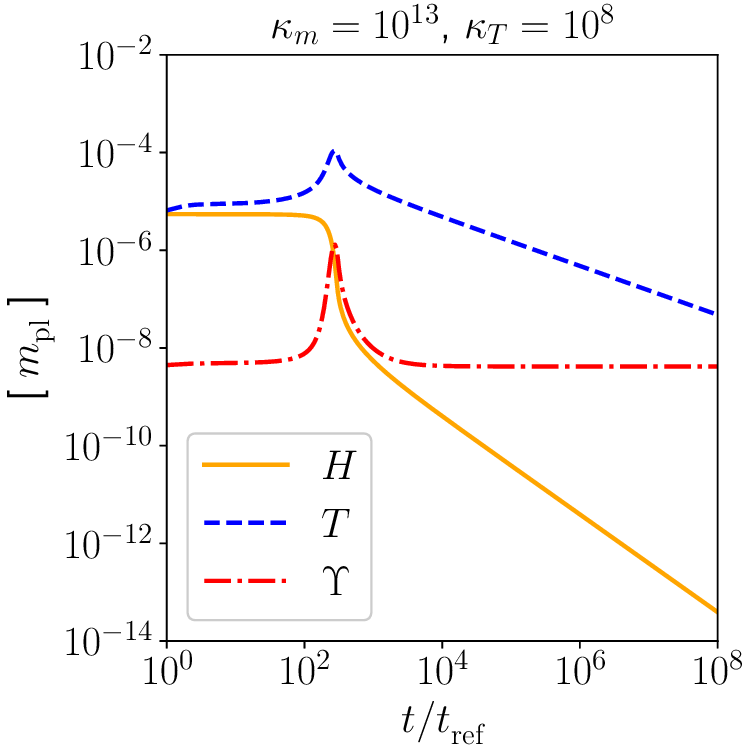}%
  \hspace{0.2cm}%
  \epsfysize=5.0cm\epsfbox{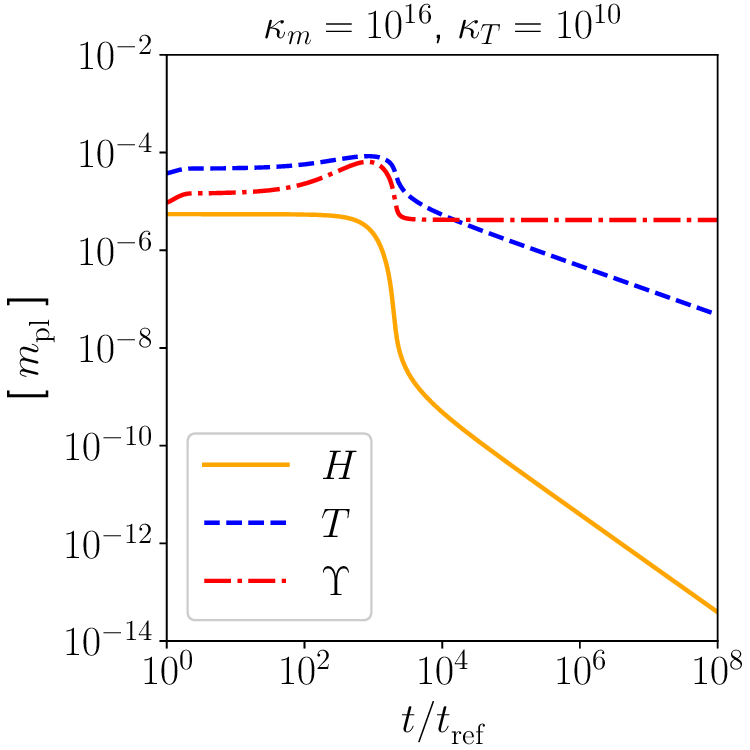}%
 }
 
 \caption[a]{\small
   Left: keeping the inflaton potential (cf.\ \eqs\nr{V}, \nr{params}) 
   and the initial conditions fixed (cf.\ \eqs\nr{init}, \nr{T_ref}), 
   the contours indicate the time $t^{ }_1$ 
   (in units of $t^{ }_\rmii{ref}$ from \eq\nr{t_ref})
   at which the ``pivot scale'', 
   $k^{ }_* / a^{ }_0 = 0.05\,\mbox{Mpc}^{-1}_{ }$, 
   reaches the value 
   $k^{ }_*/(aH) = 10^2_{ }$, 
   as a function of the parameters $\kappa^{ }_m$
   and $\kappa^{ }_\iT$ determining
   the friction (cf.\ \eq\nr{Upsilon_ansatz}).
   The perturbations are initialized at this moment, 
   cf.\ \eqs\nr{initial_dot_R_k} and \nr{initial_R_k}.  
   Middle and right: 
   the background solution from \eqs\nr{bg_varphi}--\nr{bg_H}  
   for 
   $\kappa^{ }_m = 10^{13}_{ }$, 
   $\kappa^{ }_\iT = 10^{8}_{ }$ and 
   $\kappa^{ }_m = 10^{16}_{ }$, 
   $\kappa^{ }_\iT = 10^{10}_{ }$, 
   respectively. 
   The first choice corresponds to a weak regime, 
   the second to a strong 
   regime of warm inflation, in the sense 
   defined under point~(i) in \se\ref{ss:scales}. 
   The curvature perturbations corresponding to 
   these solutions
   are shown in \fig\ref{fig:cases_curvature}.
   }
 
 \la{fig:cases_background}
\end{figure}
%%%%%%%%%%%%%%%%%%%%%%%%%%%%%%%%%%%%%%%%%%%%%%%%%%%%%%%%%%%%%%%%%%%%%%%%%%%

In order to consider physically commensurate values of $k$, 
we match the solution to a late universe. This means that we
consider a time at which inflation has ended, 
$\varphi$ has ceased to influence the background 
solution ($\Upsilon \gg H$), and the
universe undergoes radiation-dominated expansion. 
As a convenient choice, we adopt 
the moment $T = T^{ }_e \equiv 10^{-12}_{ }\mpl^{ }$, 
which is reached at the time 
$
 t^{ }_e \approx 0.301 \mpl^{ }/(\sqrt{g^{ }_*}\hspace*{0.5mm}T_e^2)
$, 
with $g^{ }_*$ defined by \eq\nr{g_star}. 
By making use
of Standard Model thermodynamics~\cite{eos15}, we can estimate 
that the scale factor increases by $a^{ }_\now/a^{ }_e \approx e^{46.5}_{ }$
from the chosen $T^{ }_e$
until today. The standard ``pivot scale'' for CMB observations today
is $k^{ }_*/a^{ }_\now = 0.05\, \mbox{Mpc}^{-1}_{ }$. So, we are 
interested in modes which have 
$
 k^{ }_*/ a^{ }_e = 
 (k^{ }_*/ a^{ }_\now) (a^{ }_\now / a^{ }_e) = 
 0.05\,\mbox{Mpc}^{-1} \, e^{46.5}_{ }
 \approx
 4.10 \times 10^{-39}_{ }\,\mpl^{ }
$.
Having determined the time history of the background
solution, we can backtrack from $t^{ }_e$ 
until a time $t^{ }_1$ at which 
\be
 \frac{ k^{ }_* }{ a^{ }_1 H^{ }_1 }
 \; = \; 
 \underbrace{
 \frac{k^{ }_*}{a^{ }_e \mpl^{ }} 
 }_{ 
 4.10 \times 10^{-39}_{ }
 }
 \underbrace{
 \frac{a^{ }_e}{ a^{ }_1} 
 }_{ 
 e^{N^{ }_e - N^{ }_1 }_{ }
 }
 \frac{\mpl^{ }}{H^{ }_1}
 \; \equiv \;
 10^2_{ } 
 \;. \la{t_1}
\ee
This is the moment at which we start evolving the density
perturbations (cf.\ \se\ref{ss:initial}). 
The value of $t^{ }_1$, in units of
$t^{ }_\rmi{ref}$ from \eq\nr{t_ref},  is plotted in the plane
$(\kappa^{ }_m,\kappa^{ }_\iT)$ in \fig\ref{fig:cases_background}(left).

%%%%%%%%%%%%%%%%%%%%%%%%%%%%%%%%% FIGURE %%%%%%%%%%%%%%%%%%%%%%%%%%%%%%%%%
\begin{figure}[t]

 \hspace*{-0.6cm}
 \centerline{%
  \epsfysize=7.2cm\epsfbox{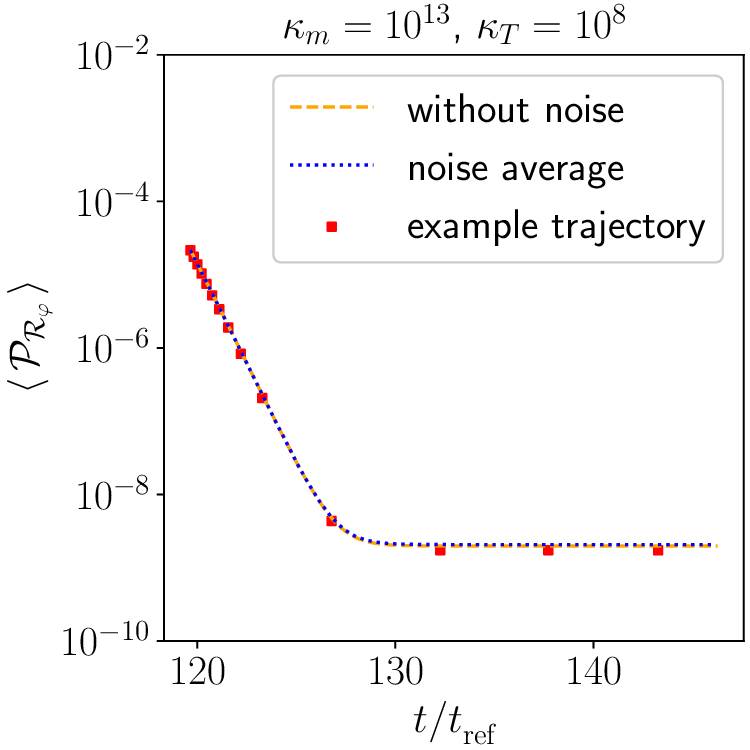}%
  \hspace{0.4cm}%
  \epsfysize=7.2cm\epsfbox{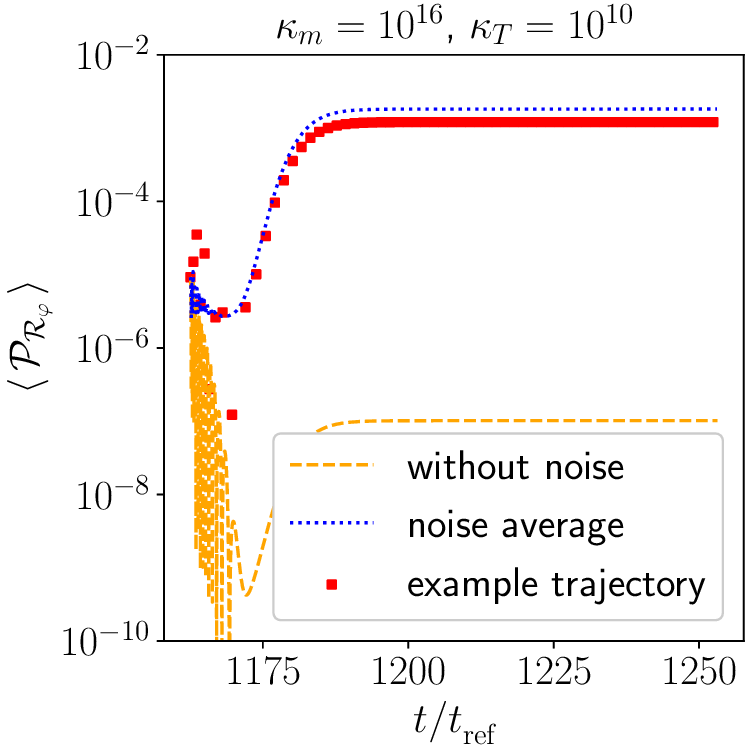}%
 }
 
 \caption[a]{\small
   The curvature perturbations corresponding to the 
   background solutions shown in \fig\ref{fig:cases_background}.
   The evolution was started at 
   $k^{ }_*/(aH) = 10^2_{ }$. 
   The dashed orange curve is a solution of 
   \eq\nr{matrix_formalism} but without noise, 
   obtained by setting
   {\tt tol}\;$=10^{-3}_{ }$ in \eq\nr{tolerance}. 
   In the dotted blue curve, the noise average has
   been included. The red squares show an example 
   trajectory from the stochastic \eq\nr{ito_1}
   (for
   better visibility, 
   only a subset of points are shown). 
   For the parameters at left, the noise
   has no visible effect. For the parameters at right, 
   the noiseless 
   solution undergoes acoustic oscillations
   while decaying.
   The noise 
   compensates for the damping by 
   generating new fluctuations.
   Around the time of the horizon exit, the 
   curvature perturbations $\R^{ }_v$ and $\R^{ }_\iT$
   drag $\R^{ }_\varphi$ to a large common value.  
   }
 
 \la{fig:cases_curvature}
\end{figure}
%%%%%%%%%%%%%%%%%%%%%%%%%%%%%%%%%%%%%%%%%%%%%%%%%%%%%%%%%%%%%%%%%%%%%%%%%%%

%%%%%%%%%%%%%%%%%%%%%%%%%%%%%%%%% FIGURE %%%%%%%%%%%%%%%%%%%%%%%%%%%%%%%%%
\begin{figure}[t]

 \hspace*{-0.6cm}
 \centerline{%
  \epsfysize=7.2cm\epsfbox{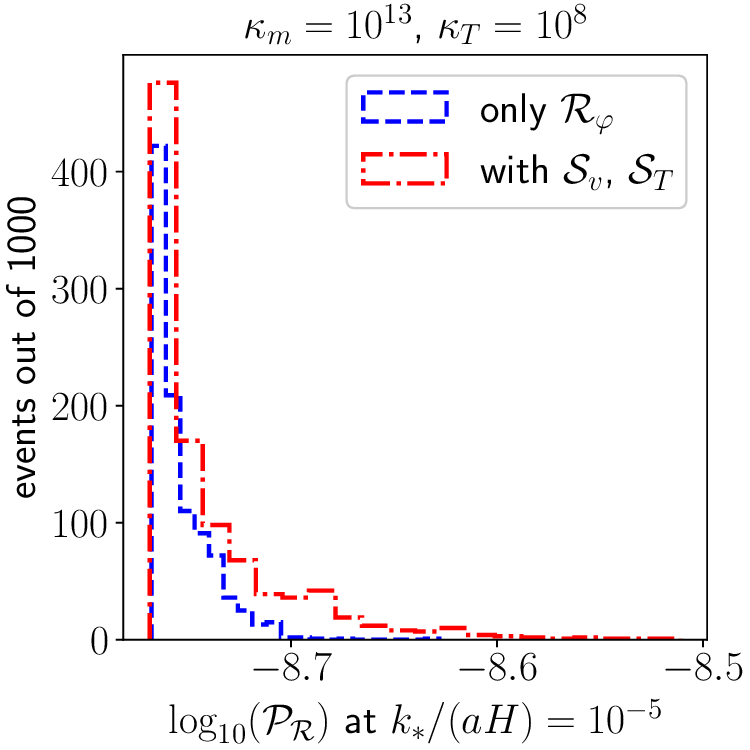}%
  \hspace{0.4cm}%
  \epsfysize=7.2cm\epsfbox{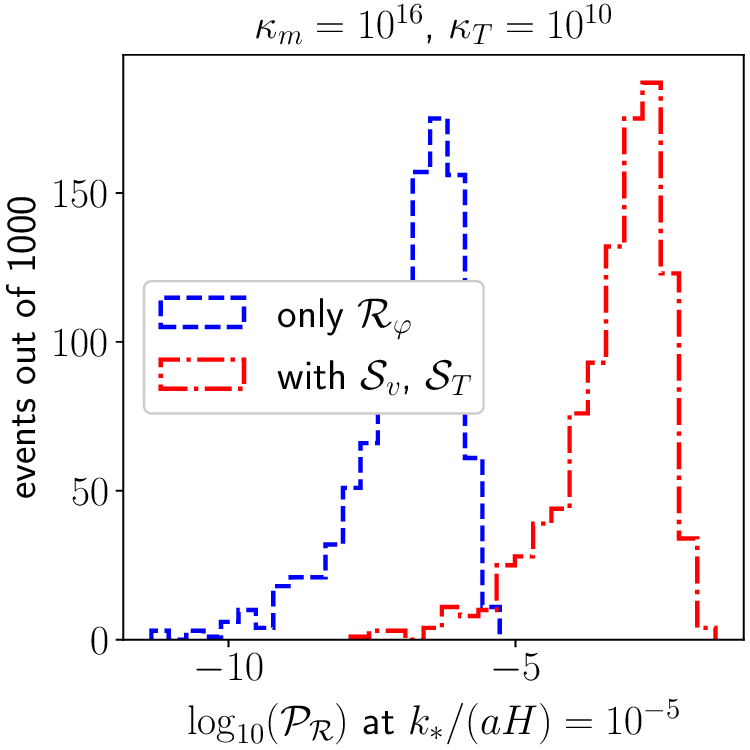}%
 }
  
 \caption[a]{\small
   The probability distributions of
   the curvature power spectra for 
   the two benchmarks
   from \fig\ref{fig:cases_curvature},
   from the weak (left) and strong (right) regime
   (the individual bins are broader for a broader distribution, 
   so the areas do not look the same). 
   The skewed distribution on the right resembles the one
   found in ref.~\cite{ballesteros}, where its shape and
   relation to $\log^{ }_{10}\langle\P^{ }_\R\rangle$ were also determined.
   }
 
 \la{fig:histograms}
\end{figure}
%%%%%%%%%%%%%%%%%%%%%%%%%%%%%%%%%%%%%%%%%%%%%%%%%%%%%%%%%%%%%%%%%%%%%%%%%%%

Numerical solutions of the background evolution 
for representative choices of 
$\kappa^{ }_m$ and $\kappa^{ }_\iT$ are shown in 
\fig\ref{fig:cases_background}(middle,right).
The time evolutions of the 
corresponding curvature power spectra are illustrated
in \fig\ref{fig:cases_curvature}. 
We show three variants: 
a solution for the averaged $\langle \P^{ }_{\R_\varphi} \rangle$
from \eq\nr{matrix_formalism} but without noise;
a solution of \eq\nr{matrix_formalism} with noise included; 
and one trajectory from a solution of the full stochastic
\eq\nr{ito_1}.  
The probability distributions from a 
statistics of $10^3_{ }$ trajectories
obtained from \eq\nr{ito_1} 
are shown in \fig\ref{fig:histograms}.

It is appropriate to remark that if $T \gg H$, then $k/(a T)$ is
moderate at the initial time, $t = t^{ }_1$
(cf.\ \eq\nr{t_1}), 
and the noise
amplitude from \eq\nr{calN_i} is not exponentially suppressed. 
However, we find that the results stay the same 
also for earlier starting times, 
say $k^{ }_*/(a^{ }_1 H^{ }_1) = 10^3_{ }$, when the exponential
suppression of the noise autocorrelator
is clearly visible. 
The reason is that 
the initial noise only excites acoustic oscillations,  
which do
not influence the value of the curvature perturbation around
its freeze-out point. 

To summarize the features observed
in \figs\ref{fig:cases_background}, \ref{fig:cases_curvature} 
and \ref{fig:histograms}, the presence of $\Upsilon,\varrho\neq 0$
influences the dynamics of the system in many ways. 
The background solution is modified, which changes
the way in which momenta redshift (cf.\ \fig\ref{fig:cases_background}(left)). 
On the side of perturbations, the original
quantum fluctuations get dissipated, if $\Upsilon > H$ before
they exit the Hubble horizon (cf.\ \fig\ref{fig:cases_curvature}). 
At the same time, new thermal fluctuations
are generated, if $T > H$ during 
the same period (cf.\ \fig\ref{fig:cases_curvature}).
An important point, visible from \ref{fig:cases_curvature}(right), 
is that the presence of $\R^{ }_v$ and $\R^{ }_\iT$
(through $\E^{ }_v$ and $\E^{ }_\iT$, cf.\ \eqs\nr{def_S_v}
and \nr{def_S_T}) has a large influence on $\R^{ }_\varphi$: 
$\E^{ }_v$ and $\E^{ }_\iT$ vanish outside of the Hubble horizon, 
and therefore $\R^{ }_v$ and $\R^{ }_\iT$ pull $\R^{ }_\varphi$
up to their common value before the solution settles to a constant. 
Effectively, this means that $\R^{ }_v$ and $\R^{ }_\iT$ from 
\eqs\nr{def_R_v} and \nr{def_R_T} determine that final value of 
$\P^{ }_\R$.
The probability distribution of $\log^{ }_{10}\P^{ }_\R$ 
is skewed (cf.\ \fig\ref{fig:histograms}), 
as elaborated upon in ref.~\cite{ballesteros}. 

%%%%%%%%%%%%%%%%%%%%%%%%%%% SECTION %%%%%%%%%%%%%%%%%%%%%%%%%%%%%%%%%%%%%%
%
\section{Can $\P^{ }_\R$ be expressed in terms of freeze-out parameters?}
\la{se:scans}

In the case of normal ``cold'' inflation, to which our setup 
reduces if $ \Upsilon,T \ll H$, it is well known that 
the out-of-horizon power spectrum, $\P^{ }_\R (k \ll aH)$, can be
approximated (within a few percent accuracy) by 
\be
 \P^\rmi{\hspace*{0.3mm}vac}_\R
 \; 
 {\equiv}
 \; 
 \frac{ H^4_{*} }{ (2\pi\dot{\bar\varphi}^{ }_{*})^2_{ } }
 \;, \la{vac}
\ee
where $f^{ }_* \equiv f(t^{ }_*)$ denotes the value of 
$f(t)$ evaluated at the time $t = t^{ }_*$ when 
the momentum scale exits the Hubble horizon, i.e.\ 
$
 k^{ } \equiv a(t^{ }_*) H(t^{ }_*)
$.
In warm inflation literature, 
it is generally assumed that the presence of fluctuations and dissipation
modifies \eq\nr{vac} through a function which depends on 
the values of the medium-induced 
quantities at the freeze-out point, normalized
to the corresponding Hubble rate, notably 
$T^{ }_* / H^{ }_*$ and $\Upsilon^{ }_* / H^{ }_*$
(cf.\ \eqs\nr{fit_G_1} and \nr{fit_G_2}).
We refer to these ratios as {\em freeze-out parameters}. 

An immediate remark is that, as is visible in \eq\nr{vac}, 
we should consider $|\dot{\bar\varphi}^{ }_*|/H_*^2$ as an additional
freeze-out parameter. 
In fact, 
$\dot{\bar\varphi}^2_{ }$ plays a prominent role
in \eqs\nr{dot_S_v} and \nr{dot_S_T}. 
In phenomenological scans, the value 
of $|\dot{\bar\varphi}^{ }_*|/H_*^2$ varies by many orders of 
magnitude (cf.\ table~\ref{table:sm} on p.~\pageref{table:sm}). 
However, the way that we
have organized the scans of the present section, 
by keeping the potential
fixed (cf.\ appendix~\ref{se:V}), implies that 
$|\dot{\bar\varphi}^{ }_*|/H_*^2$ only varies by  
$\rmO(1)$ (cf.\ \fig\ref{fig:scans_bg}). This simplifies our
task of identifying possible functional dependences.  

%%%%%%%%%%%%%%%%%%%%%%%%%%%%%%%%% FIGURE %%%%%%%%%%%%%%%%%%%%%%%%%%%%%%%%%
\begin{figure}[t]

 \hspace*{-0.3cm}
 \centerline{%
  \epsfysize=5.0cm\epsfbox{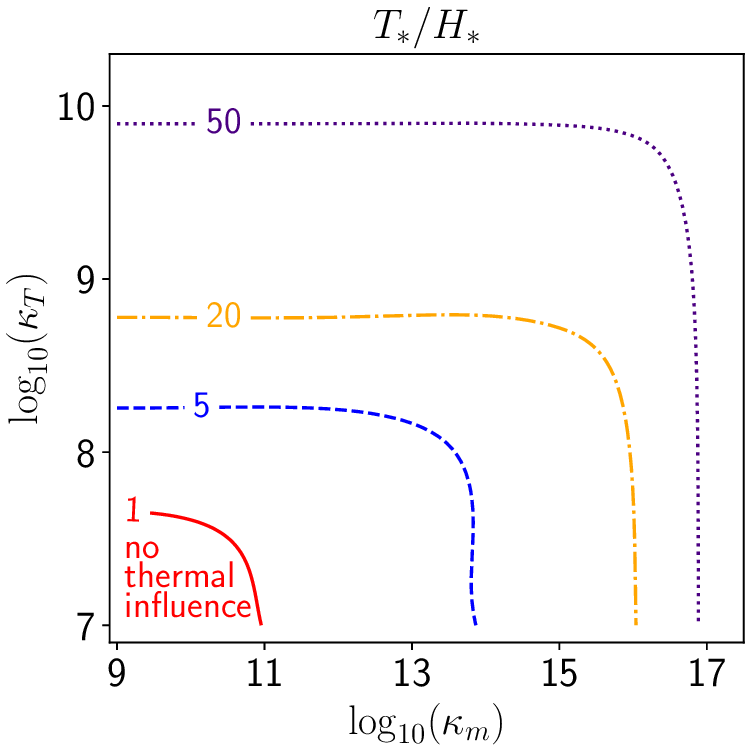}%
  \hspace{0.2cm}%
  \epsfysize=5.0cm\epsfbox{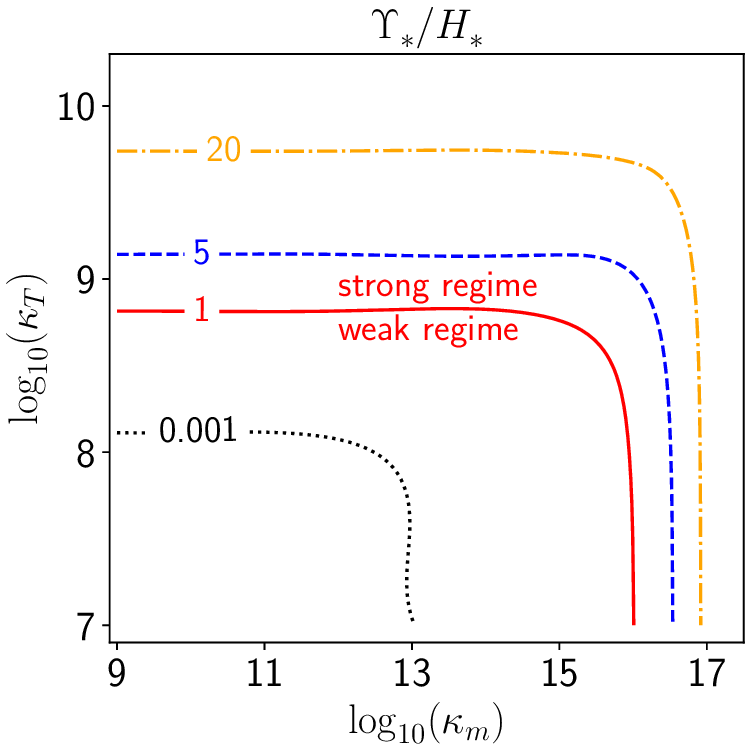}%
  \hspace{0.2cm}%
  \epsfysize=5.0cm\epsfbox{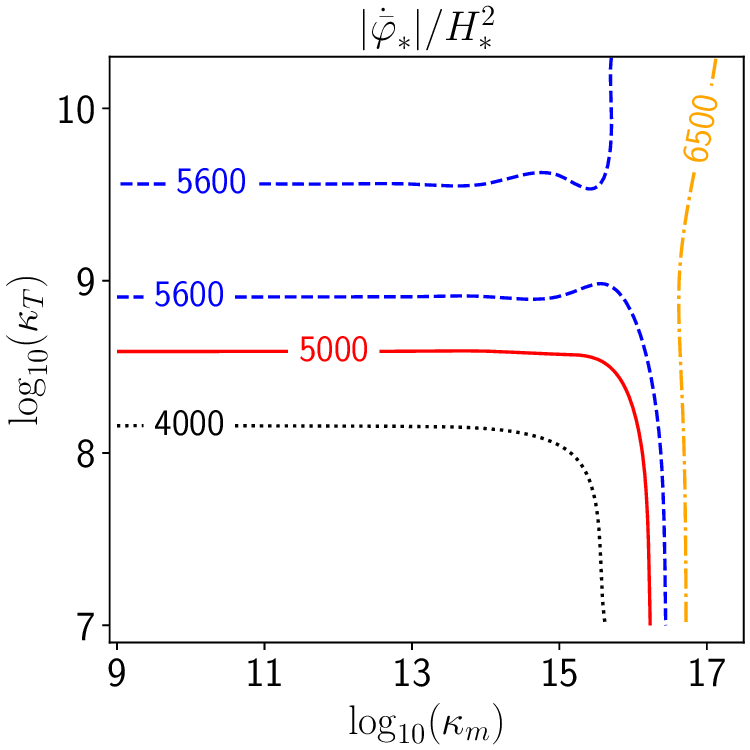}%
 }
 
 \caption[a]{\small
   The main physical characteristics 
   of the background solution 
   at the ``freeze-out point'', $k^{ }_*/(a H) \equiv 1$,  
   as a function of the parameters $\kappa^{ }_m$
   and $\kappa^{ }_\iT$ parametrizing $\Upsilon$ (cf.\ \eq\nr{Upsilon_ansatz}).
   Left: the temperature $T$, normalized to 
   the Hubble rate~$H$.
   Middle: like at left, but for the friction coefficient~$\Upsilon$. 
   The ``strong'' and ``weak'' regime, as well as
   the corner where the thermal plasma has no influence, 
   refer to item~(i) in \se\ref{ss:scales}.
   Right: the inflaton derivative, $\dot{\bar\varphi}$, 
   normalized to $H^2_{ }$. Even if the value does not vary much, 
   the variation has a non-trivial shape, which correlates
   in an intriguing way with curvature perturbations (cf.\ the text).
   }
 
 \la{fig:scans_bg}
\end{figure}
%%%%%%%%%%%%%%%%%%%%%%%%%%%%%%%%%%%%%%%%%%%%%%%%%%%%%%%%%%%%%%%%%%%%%%%%%%%

The values of the freeze-out parameters
$ T^{ }_*/H^{ }_*  $, 
$ \Upsilon^{ }_*/H^{ }_*  $ and 
$|\dot{\bar\varphi}^{ }_*| / H_*^2$ are
shown in \fig\ref{fig:scans_bg}. 
The similar shapes of 
the contours of constant 
$ T^{ }_*/H^{ }_*  $ and
$ \Upsilon^{ }_*/H^{ }_*  $ 
suggest that these parameters are strongly correlated.
The contours of constant 
$|\dot{\bar\varphi}^{ }_*| / H_*^2$ show a 
different shape, particularly in the strong regime, 
however the variation is mild. 

%%%%%%%%%%%%%%%%%%%%%%%%%%%%%%%%% FIGURE %%%%%%%%%%%%%%%%%%%%%%%%%%%%%%%%%
\begin{figure}[t]

 \hspace*{-0.1cm}
 \centerline{%
  \epsfysize=7.4cm\epsfbox{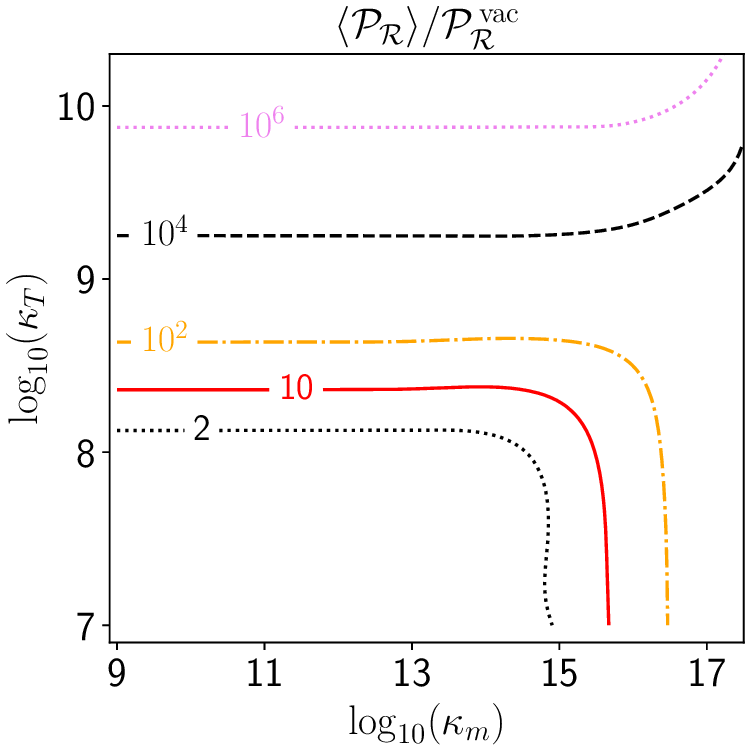}%
  \hspace{0.2cm}%
  \epsfysize=7.4cm\epsfbox{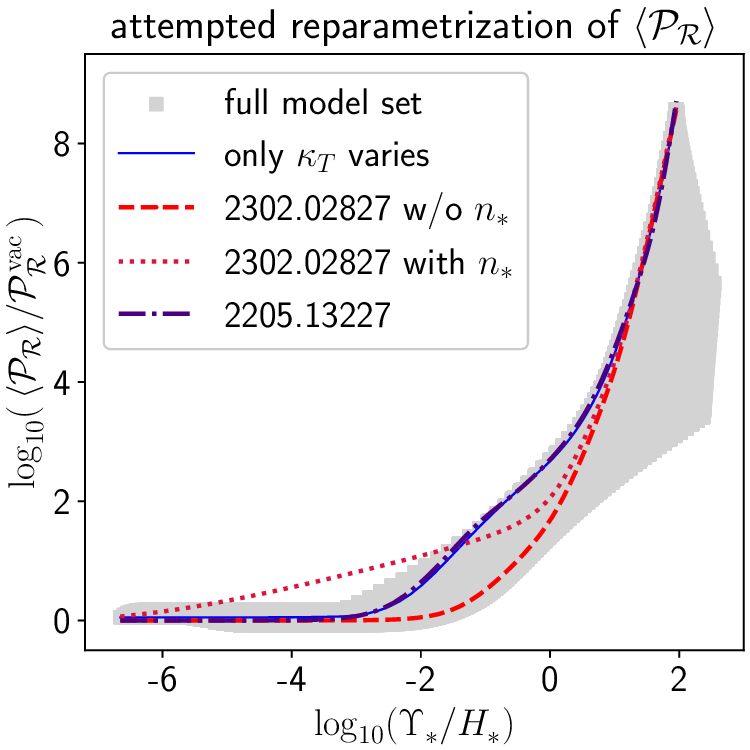}%
 }
 
 \caption[a]{\small
   Left: 
   $\langle\P^{ }_\R\rangle / \P^\rmi{\hspace*{0.3mm}vac}_\R $
   outside of 
   the Hubble horizon, $k^{ }_*/(a H) \equiv 10^{-5}_{ }$,  
   as a function of $\kappa^{ }_m$
   and $\kappa^{ }_\iT$ parametrizing $\Upsilon$ (cf.\ \eq\nr{Upsilon_ansatz}).
   Right: 
   $\langle\P^{ }_\R\rangle / \P^\rmi{\hspace*{0.3mm}vac}_\R $
   replotted versus the freeze-out parameter $\Upsilon^{ }_*/H^{ }_*$
   from \fig\ref{fig:scans_bg}(middle).
   We also compare a specific model
   (``only $\kappa^{ }_\iT$ varies'', implying $\Upsilon\sim T^3_{ }$)
   with fit forms from the literature (cf.\ the text).
   The grey band shows that $\langle\P^{ }_\R\rangle$ is {\em not} 
   a single-valued function of $Q^{ }_* = \Upsilon^{ }_*/(3 H^{ }_*)$,
   but displays a large spread as $Q^{ }_*$ increases. 
   }
 
 \la{fig:fit}
\end{figure}
%%%%%%%%%%%%%%%%%%%%%%%%%%%%%%%%%%%%%%%%%%%%%%%%%%%%%%%%%%%%%%%%%%%%%%%%%%%

The key question is whether $\P^{ }_\R$ respects the shapes of 
the curves of constant freeze-out parameters. 
The value 
$\log^{ }_{10}( \langle \P^{ }_\R \rangle /
 \P^\rmi{\hspace*{0.3mm}vac}_\R )$ 
is plotted in \fig\ref{fig:fit}(left).
Clearly, the curves 
do {\em not} display the same geometry as  
the contours of constant 
$T^{ }_*/H^{ }_*$ and $\Upsilon^{}_*/H^{ }_*$
in \fig\ref{fig:scans_bg}. They resemble more 
those of constant $|\dot{\bar\varphi}^{ }_*| / H_*^2$, 
even if only qualitatively. 

% We remark that in warm inflation literature, 
% the value of $Q \equiv \Upsilon / (3 H)$ is frequently treated as
% an input parameter, kept fixed as a function of time. Then it 
% is not possible to distinguish between its roles as a model parameter,
% or a model-independent freeze-out parameter. 

To be more concrete, 
we denote $Q^{ }_* \equiv \Upsilon^{ }_* / (3 H^{ }_*)$,
and recall two ans\"atze for $\langle \P^{ }_\R \rangle$
from warm inflation literature,
\ba
 \langle \P^{ }_\R \rangle
 & 
 \overset{\rmii{\cite{warm1},\,\eq(4.26)}}{
 \underset{ }{\equiv}}
 & 
 \overbrace{
 \frac{ H^4_{*} }{ (2\pi\dot{\bar\varphi}^{ }_{*})^2_{ } }
 }^{\equiv\;\P^\rmii{\hspace*{0.3mm}vac}_\R }
 \; 
 \biggl(\, 
 1 + 
 \overbrace{
 \frac{2 \hspace*{0.3mm}\theta^{ }_* }{e^{H^{ }_*/T^{ }_*}_{ } - 1}
 }^{ \;\equiv\; 2 \hspace*{0.3mm} n^{ }_* }
 + \frac{2\sqrt{3} \pi Q^{ }_*}{\sqrt{3 + 4\pi Q^{ }_*}}
   \, \frac{T^{ }_*}{H^{ }_*}
 \,\biggr)
 \, {\G}(Q^{ }_*)
 \la{fit_G_1}
 \\
%%%%%%%%%
 & 
 \overset{\rmii{\cite{mehrdad}, \eq(49)}}{
 \underset{ }{\equiv}}
 & 
 \frac{ H^4_{*} }{ (2\pi\dot{\bar\varphi}^{ }_{*})^2_{ } }
 \;
 \biggl[\, 
 1 + 
 \frac{12 Q^{ }_* T^{ }_* \, F^{ }_2\bigl( 3 Q^{ }_*  \bigr) }{H_*^{ }}
 \,\biggr]
 \;, \la{fit_G_2}
\ea
where, 
according to ref.~\cite{warm1}, 
$\theta^{ }_* = 1$ % in \eq\nr{fit_G_1} 
if $\varphi$ is thermalized, 
otherwise $\theta^{ }_* = 0$.\footnote{%
 The inclusion of $2 n^{ }_*$ was 
 questioned in ref.~\cite{ballesteros}. The origin of this factor 
 can be seen in \eq\nr{P_Q_k_thermal}.  
 As shown in \eq\nr{res_Omega}, when we match quantum mechanics 
 to the classical Langevin description, the Bose distribution 
 appears through the noise autocorrelator. The linear~$T^{ }_*$
 dependence, visible in \eqs\nr{fit_G_1} and \nr{fit_G_2}, 
 is a special kinematic limit
 of $n^{ }_\rmiii{B}$, cf.\ \eq\nr{res_Omega_ir}.  
 In our approach, $n^{ }_\rmiii{B}$ appears nowhere else than the
 noise autocorrelator, and the thermalization of the low-momentum
 modes of $\varphi$ is determined by the dynamics, 
 rather than having to be put in by hand. However, 
 $\Upsilon$ is momentum dependent, and the equilibration of 
 the high-momentum modes of $\varphi$ remains a separate topic.  
 }
Obviously, \eqs\nr{fit_G_1} and
\nr{fit_G_2} can always be postulated if the result depends
on a single parameter, with the function ${\G}$ or $F^{ }_2$ 
adjusted to capture the parameter dependence. The non-trivial question
is if such a representation could be {\em model-independent}, 
like the vacuum limit in \eq\nr{vac}, which
applies to any potential $V$ and any momentum $k$
(whether close to the pivot scale or not, 
and whether matched to CMB data or not).
Let us anticipate that we find the answer to be negative, 
and do {\em not} make use of 
\eq\nr{fit_G_1} or \nr{fit_G_2} in our own analysis, 
however we compare with them for a particular 
single-parameter family of variations 
(cf.\ \fig\ref{fig:fit}(right)).

%%%%%%%%%%%%%%%%%%%%%%%%%%%%%%%%% FIGURE %%%%%%%%%%%%%%%%%%%%%%%%%%%%%%%%%
\begin{figure}[t]

 \hspace*{-0.1cm}
 \centerline{%
  \epsfysize=7.4cm\epsfbox{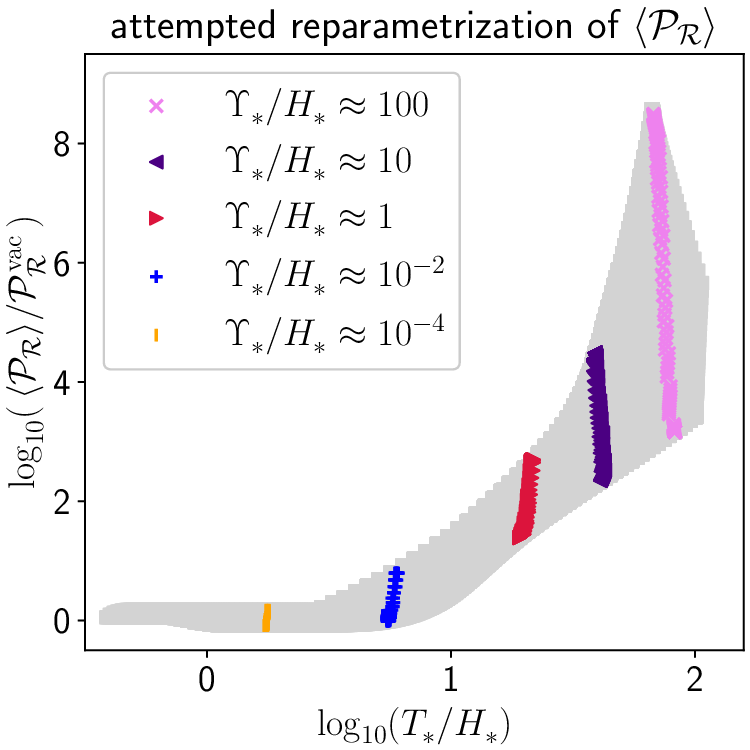}%
  \hspace{0.2cm}%
  \epsfysize=7.4cm\epsfbox{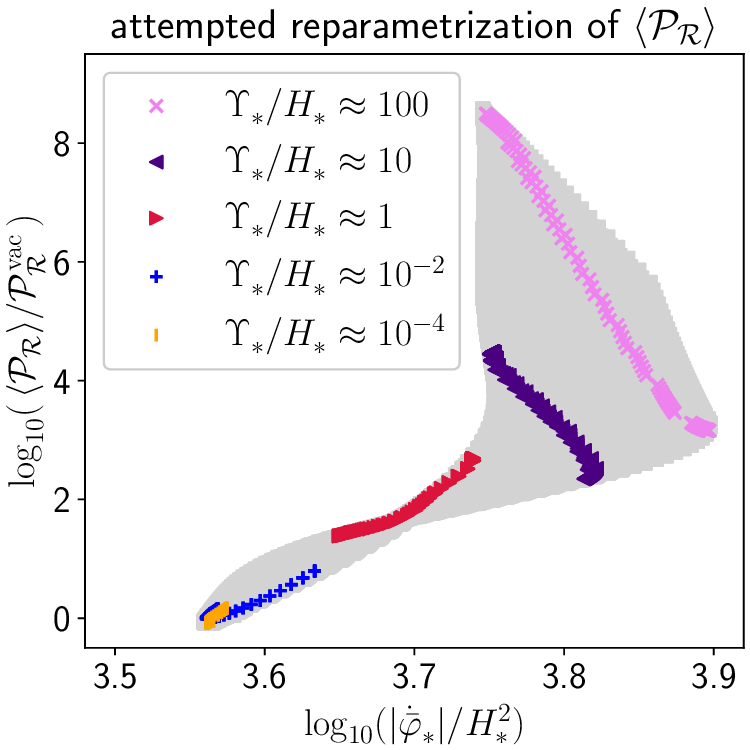}%
 }
 
 \caption[a]{\small
   $\langle\P^{ }_\R\rangle$ from \fig\ref{fig:fit}(left), 
   replotted versus the freeze-out parameters $T^{ }_*/H^{ }_*$
   and $|\dot{\bar\varphi}^{ }_*|/H_*^2$
   from \fig\ref{fig:scans_bg}.
   The left panel shows that the 
   multi-valuedness of $\langle\P^{ }_\R\rangle
   / \P^\rmi{\hspace*{0.3mm}vac}_\R$ 
   is {\em not} efficiently resolved by including
   $T^{ }_*/H^{ }_*$ as an additional freeze-out parameter: 
   the chosen subsets are almost vertical in $T^{ }_*/H^{ }_*$. 
   The right panel shows that results 
   vary noticeably with $|\dot{\bar\varphi}^{ }_*|/H_*^2$, however this
   is partly an optical illusion, as we have zoomed into a very narrow
   range of values. Nevertheless, it is interesting that the 
   sign of the correlation switches between 
   $\Upsilon^{ }_*/H^{ }_* = 1$ and $10$, when 
   the radiation energy density overtakes the inflaton kinetic energy, 
   $e^{ }_r > \dot{\bar\varphi}^2_{ } / 2$. 
   
   }
 
 \la{fig:repara}
\end{figure}
%%%%%%%%%%%%%%%%%%%%%%%%%%%%%%%%%%%%%%%%%%%%%%%%%%%%%%%%%%%%%%%%%%%%%%%%%%%

Now, physical intuition might 
suggest that the final result {\em should be} a function 
of the freeze-out parameters. The simplest way
to say this is that a classical thermal state has no memory. All 
its properties should be fixed by a few macroscopic control 
parameters, as they appear in the evolution equations. 
Evidence for a certain degree of 
universality was put forward 
in ref.~\cite{freese}. 
It was demonstrated 
that ${\G}$ is independent of the 
$\varphi$-dependence of $\Upsilon$, the inflaton potential $V$, 
and the number of $e$-folds (or the momentum $k$). Also, it was
stated that it does not matter if $Q$ is evolved or treated as 
constant (though most of the runs kept it time-independent). 
At the same time, significant non-universality was observed
between models with different $T$-dependences of $Q$. 

We can scrutinize numerically the universality of 
the curvature power spectrum. 
Our tool is that the scans in \se\ref{se:scans} span a two-parameter set of
models. We can then replace the original parameters 
% ($\kappa^{ }_m$, $\kappa^{ }_\iT$) 
through the would-be freeze-out parameters. We replot the scan
results from \fig\ref{fig:fit}(left) as a function of various 
freeze-out parameters in \figs\ref{fig:fit}(right) and \ref{fig:repara}. 

As a first test, 
the solid blue line in \fig\ref{fig:fit}(right) 
shows the result obtained if we keep~$\kappa^{ }_m$ fixed
to its minimum, $\kappa^{ }_m = 10^9_{ }$, 
and only let $\kappa^{ }_\iT$ vary. Then, 
effectively, $\Upsilon \sim T^3_{ }$, a case that has been studied
in the literature (cf.,\ e.g.,\ ref.~\cite{mwi}). 
We show a comparison with employing \eq(4.27) of ref.~\cite{warm1}
in \eq\nr{fit_G_1} 
[``2302.02827''], 
and \eq(47) of ref.~\cite{mehrdad} in \eq\nr{fit_G_2}
[``2205.13227'']. 
Our results agree well with ref.~\cite{mehrdad}. 
As can be inferred from ref.~\cite{ramos}, 
the difference with
respect to ref.~\cite{warm1} 
is likely due to 
the omission of the stochastic noise from 
\eq\nr{dot_S_T}, 
and the fact that 
the fit only considered the regime $\Upsilon^{ }_*/H^{ }_* \gg 1$, 
even though it is regularly applied also at $\Upsilon^{ }_*/H^{ }_* \ll 1$.

Our main diagnostics is shown in \fig\ref{fig:repara}.
In the left panel, 
we highlight subsets with a specific $\Upsilon^{ }_*/H^{ }_*$, 
as a function of $T^{ }_*/H^{ }_*$. However, the sets remain practically
vertical: 
trajectories of 
constant $\Upsilon^{ }_*/H^{ }_*$  correspond
to constant $T^{ }_*/H^{ }_*$, 
so that $T^{ }_*/H^{ }_*$ has no resolving power.  
% This conclusion is the same 
% (in fact, even stronger) if we factor out a linear 
% $T^{ }_*/H^{ }_*$ dependence, like in \eqs\nr{fit_G_1} 
% and \nr{fit_G_2}. 

In contrast, the right panel of \fig\ref{fig:repara} shows 
the dependence on $|\dot{\bar\varphi}^{ }_*|/H_*^2$, 
revealing a correlation. Notably, in the weak regime, 
$\langle\P^{ }_\R\rangle / \P^\rmi{\hspace*{0.3mm}vac}_\R$ grows 
with $|\dot{\bar\varphi}^{ }_*|/H_*^2$, 
whereas in the strong regime, it decreases. 
However, in both cases, 
$|\dot{\bar\varphi}^{ }_*|/H_*^2$ only varies very little, 
so we cannot represent an orders-of-magnitude change of 
$\langle\P^{ }_\R\rangle / \P^\rmi{\hspace*{0.3mm}vac}_\R$ 
in any natural way (e.g.\ a power). 

%%%%%%%%%%%%%%%%%%%%%%%%%%%%%%%%% FIGURE %%%%%%%%%%%%%%%%%%%%%%%%%%%%%%%%%
\begin{figure}[t]

 \hspace*{-0.1cm}
 \centerline{%
  \epsfysize=7.4cm\epsfbox{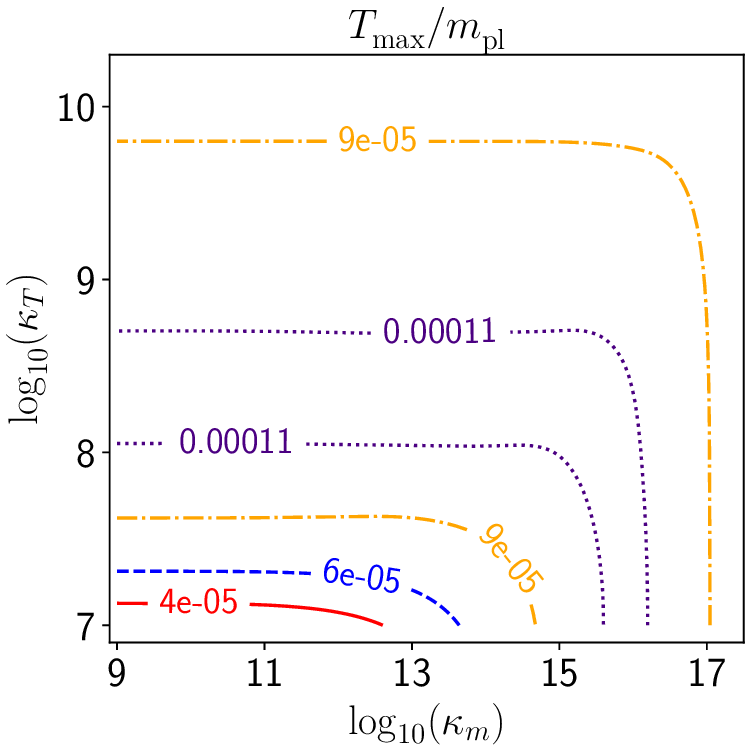}%
  \hspace{0.2cm}%
  \epsfysize=7.4cm\epsfbox{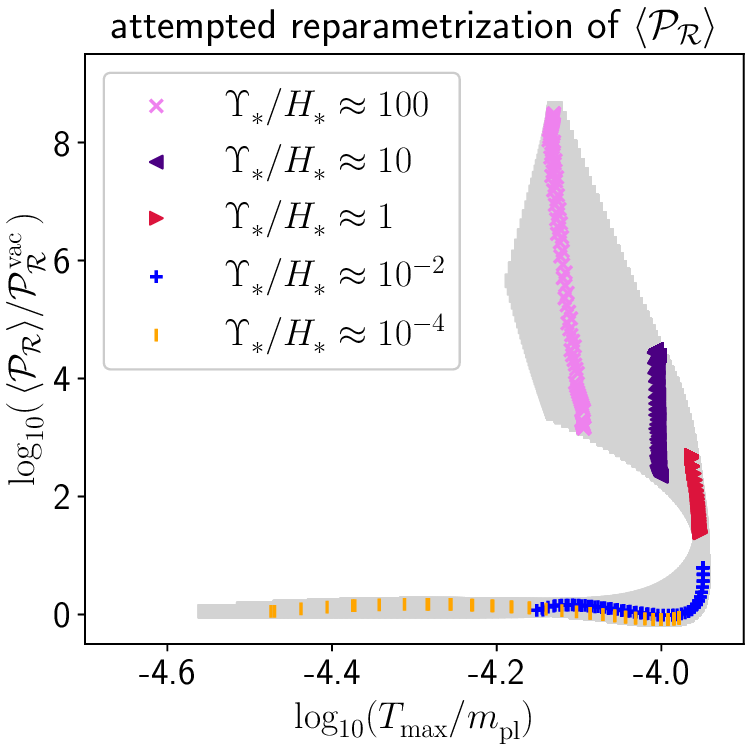}%
 }
 
 \caption[a]{\small
   Left: 
   the maximal temperature reached during the reheating history, 
   as a function of $\kappa^{ }_m$
   and $\kappa^{ }_\iT$ parametrizing $\Upsilon$ (cf.\ \eq\nr{Upsilon_ansatz}).
   Even though $\Upsilon$ varies by many orders of magnitude, 
   $T^{ }_\rmii{max}$ only varies modestly. 
   Right: 
   an attempted reparametrization of 
   $\langle\P^{ }_\R\rangle$ from \fig\ref{fig:fit}(left), 
   versus $T^{ }_\rmii{max}$.
   We observe that in the weak regime, 
   $\langle\P^{ }_\R\rangle
   / \P^\rmi{\hspace*{0.3mm}vac}_\R$ 
   is practically independent of $T^{ }_\rmii{max}$, 
   whereas in the strong regime, 
   the value of $\Upsilon^{ }_*/ H^{ }_*$
   effectively fixes that of $T^{ }_\rmii{max}/\mpl^{ }$, 
   so that the latter {\em cannot} resolve the model dependence
   (vertical variation) in a natural way.  
   }
 
 \la{fig:Tmax}
\end{figure}
%%%%%%%%%%%%%%%%%%%%%%%%%%%%%%%%%%%%%%%%%%%%%%%%%%%%%%%%%%%%%%%%%%%%%%%%%%%

To summarize, a large model dependence remains present in the power 
spectrum, if we try to capture the results by only making use of the 
freeze-out parameters $T^{ }_*/H^{ }_*$ and $\Upsilon^{ }_*/H^{ }_*$. 
We have seen indications, however, that additional freeze-out parameters, 
perhaps involving the energy densities 
$\dot{\bar\varphi}^2_*/2$ and $e^{ }_r$, 
could help to capture the model dependence. Unfortunately, 
given the complicated structure
of the coupled evolution equations \nr{dot_R_k}--\nr{dot_S_T}, we have 
not been able to identify analytically the precise 
quantities that could take this role. 

\vspace*{3mm}

To end this section, we show a different correlation, not with 
freeze-out parameters but with the maximal temperature reached
during the evolution ($T^{ }_\rmi{max}$). Its value in the plane
of $\kappa^{ }_m$ and $\kappa^{ }_\iT$
is plotted in \fig\ref{fig:Tmax}(left); 
compared with $T^{ }_*/H^{ }_*$ in \fig\ref{fig:scans_bg}(left),
the shape is quite different. 
An attempted reparametrization of 
$\langle\P^{ }_\R\rangle / \P^\rmi{\hspace*{0.3mm}vac}_\R$ 
in terms of $T^{ }_\rmi{max}/\mpl^{ }$ is illustrated
in \fig\ref{fig:Tmax}(right).
Even though we again observe a geometrically interesting correlation, 
and a very clear difference between the weak and strong regimes, the plot
does not suggest any natural way to resolve the degeneracy
in both regimes. 
It is also appropriate to remark that in some cases, 
notably in the model considered in \se\ref{se:sm}, the temperature
does not peak, but is rather a slowly decreasing function
during the reheating epoch. 

%%%%%%%%%%%%%%%%%%%%%%%%%%% SECTION %%%%%%%%%%%%%%%%%%%%%%%%%%%%%%%%%%%%%%
%
\section{Testing a Standard Model embedded warm inflation scenario}
\la{se:sm}

%%%%%%%%%%%%%%%%%%%%% TABLE %%%%%%%%%%%%%%%%%%%%%%%%%%%%%%%%%%%%%
%
\begin{table}[t]

\vspace*{-3mm}

{\fontsize{8pt}{10pt}\selectfont
$$
\begin{array}{|ccc|cccccccc|} % {llllllllll} %  {rrrrrrrrrr}
%%%%%%%%%%%%%%%
  \hline
  & & & & & & & & & & \\[-3mm]
  \multicolumn{3}{|c|}{ \mbox{original~input~[2503.18829v2]} } 
  & 
  \multicolumn{8}{|c|}{ \mbox{our~output} } 
 \\[1mm]
%%%%%%%%%%%%%%%
  \lambda & 
  f^{ }_a \; [10^{12}_{ }\;\mbox{GeV}] &
  \alphas^{ } &
  t^{ }_1 / t^{ }_\rmii{ref} & 
  \bar\varphi^{ }_1/\mpl^{ } & 
  T^{ }_* / H^{ }_* & 
  Q^{ }_* & % \Upsilon^{ }_* / 3H^{ }_* &
  |\dot{\bar\varphi}|^{ }_* / H^2_* &
  A^{ }_\rmii{s} &
  n^{ }_\rmii{s} &
  r 
 \\[1mm]
%%%%%%%%%%%%%%%
  \hline % 
  & & & & & & & & & & \\[-3mm]
%%%%
  10^{-21}_{ } & 
  0.146 & 
  0.0269 & 
  7417 & % t_1
  1.03 & % varphi*/mpl
  8042 & % T*/H*
  14.9 & % Ups*/H*
  1.14p8 & % phid*/HH*
  2.73m9 & % As
  0.967 & % ns 
  1.46m11   % r
   \\[1mm]
%%%%
  10^{-20}_{ } & 
  0.264 & 
  0.0263 & 
  4310 & % t_1
  1.26 & % varphi*/mpl
  3660 & % T*/H*
  9.25 & % Ups*/H*
  3.01p7 & % phid*/HH*
  2.70m9 & % As
  0.969 & % ns 
  3.39m10   % r
   \\[1mm]
%%%%
  10^{-19}_{ } & 
  0.518 & 
  0.0257 & 
  2121 & % t_1
  1.67 & % varphi*/mpl
  1527 & % T*/H*
  4.85 & % Ups*/H*
  7.22p6 & % phid*/HH*
  2.07m9 & % As
  0.968 & % ns 
  1.35m8   % r
   \\[1mm]
%%%%
  10^{-18}_{ } & 
  1.05 & 
  0.0251 & 
  954 & % t_1
  2.27 & % varphi*/mpl
  601 & % T*/H*
  2.21 & % Ups*/H*
  1.66p6 & % phid*/HH*
  1.93m9 & % As
  0.968 & % ns 
  4.93m7   % r
   \\[1mm]
%%%%
  10^{-17}_{ } & 
  2.30 & 
  0.0246 & 
  373 & % t_1
  3.18 & % varphi*/mpl
  211 & % T*/H*
  0.687 & % Ups*/H*
  3.65p5 & % phid*/HH*
  2.03m9 & % As
  0.967 & % ns 
  1.80m5   % r
   \\[1mm]
%%%%
  10^{-16}_{ } & 
  4.95 & 
  0.0243 & 
  178 & % t_1
  4.07 & % varphi*/mpl
  61.6 & % T*/H*
  0.0951 & % Ups*/H*
  8.40p4 & % phid*/HH*
  2.15m9 & % As
  0.971 & % ns 
  4.62m4   % r
   \\[1mm]
%%%%
  10^{-15}_{ } & 
  8.69 & 
  0.0242 & 
  148 & % t_1
  4.36 & % varphi*/mpl
  19.8 & % T*/H*
  0.0134 & % Ups*/H*
  2.32p4 & % phid*/HH*
  1.85m9 & % As
  0.980 & % ns 
  7.15m3   % r
   \\[1mm]
%%%%
  \hline
\end{array}
$$

%%%%%%%%%%%%%%%%%%%%%%%%%%%%%%%%%%%%%%%%%%%%%%%%%%%%%%%%%%%%%%%%%%%
\vspace*{-3mm}

$$
\begin{array}{|ccc|cccccccc|} % {llllllllll} %  {rrrrrrrrrr}
%%%%%%%%%%%%%%%
  \hline
  & & & & & & & & & & \\[-3mm]
  \multicolumn{3}{|c|}{
   \hspace*{11.2mm} \mbox{adjusted~input} \hspace*{11.2mm} } 
  & 
  \multicolumn{8}{|c|}{ \mbox{our~output} } 
 \\[1mm]
%%%%%%%%%%%%%%%
  \lambda & 
  f^{ }_a \; [10^{12}_{ }\;\mbox{GeV}] &
  \alphas^{ } &
  t^{ }_1 / t^{ }_\rmii{ref} & 
  \bar\varphi^{ }_1/\mpl^{ } & 
  T^{ }_* / H^{ }_* & 
  Q^{ }_* & % \Upsilon^{ }_* / 3H^{ }_* &
  |\dot{\bar\varphi}|^{ }_* / H^2_* &
  A^{ }_\rmii{s} &
  n^{ }_\rmii{s} &
  r 
 \\[1mm]
%%%%%%%%%%%%%%%
  \hline % 
  & & & & & & & & & & \\[-3mm]
%%%%
  10^{-21}_{ } & 
  0.153 & 
  0.0269 & 
  6897 & % t_1
  1.06 & % varphi*/mpl
  7798 & % T*/H*
  13.9 & % Ups*/H*
  1.11p8 & % phid*/HH*
  2.11m9 & % As
  0.968 & % ns 
  2.12m11   % r
   \\[1mm]
%%%%
  10^{-20}_{ } & 
  0.278 & 
  0.0263 & 
  3969 & % t_1
  1.31 & % varphi*/mpl
  3534 & % T*/H*
  8.59 & % Ups*/H*
  2.91p7 & % phid*/HH*
  2.09m9 & % As
  0.970 & % ns 
  5.00m10   % r
   \\[1mm]
%%%%
  10^{-19}_{ } & 
  0.516 & 
  0.0257 & 
  2135 & % t_1
  1.67 & % varphi*/mpl
  1531 & % T*/H*
  4.88 & % Ups*/H*
  7.24p6 & % phid*/HH*
  2.11m9 & % As
  0.969 & % ns 
  1.31m8   % r
   \\[1mm]
%%%%
  10^{-18}_{ } & 
  1.025 & 
  0.0251 & 
  994 & % t_1
  2.24 & % varphi*/mpl
  618 & % T*/H*
  2.30 & % Ups*/H*
  1.69p6 & % phid*/HH*
  2.10m9 & % As
  0.968 & % ns 
  4.27m7   % r
   \\[1mm]
%%%%
  10^{-17}_{ } & 
  2.27 & 
  0.0246 & 
  381 & % t_1
  3.15 & % varphi*/mpl
  213 & % T*/H*
  0.711 & % Ups*/H*
  3.68p5 & % phid*/HH*
  2.10m9 & % As
  0.967 & % ns 
  1.68m5  % r
   \\[1mm]
%%%%
  10^{-16}_{ } & 
  4.98 & 
  0.0243 & 
  177 & % t_1
  4.08 & % varphi*/mpl
  60.9 & % T*/H*
  0.0919 & % Ups*/H*
  8.38p4 & % phid*/HH*
  2.09m9 & % As
  0.971 & % ns 
  4.79m4   % r
   \\[1mm]
%%%%
  10^{-15}_{ } & 
  8.56 & 
  0.0242 & 
  149 & % t_1
  4.35 & % varphi*/mpl
  20.4 & % T*/H*
  0.0150 & % Ups*/H*
  2.33p4 & % phid*/HH*
  2.09m9 & % As
  0.980 & % ns 
  6.26m3   % r
   \\[1mm]
%%%%
  \hline
\end{array}
$$
}

\vspace*{-3mm}

\caption[a]{\small
 The original~\cite{sm1} (top panel) and adjusted (bottom panel)
 parameters for a scenario
 defined by the potential and friction coefficient in 
 \eqs\nr{V_sm} and \nr{Upsilon_sm}, respectively. 
 Here $t^{ }_1$ is the time at which $k^{ }_*/(aH) = 10^2_{ }$
 and we start evolving perturbations;
% (like in \fig\ref{fig:cases_background}(left)); 
 $t^{ }_\rmii{ref}$ is defined in \eq\nr{tref_sm} and agrees in order of
 magnitude with the inverse of the initial Hubble rate; 
 the freeze-out parameters are  
 evaluated when $k^{ }_*/(aH) = 1$;
% correspond to those in \fig\ref{fig:scans_bg}; 
 and $A^{ }_\rmi{s}$, $n^{ }_\rmi{s}$, $r$
% , 
% obtained from \eqs\nr{As}, \nr{ns}, and \nr{r}, respectively, 
 are evaluated when $k^{ }_*/(aH) = 10^{-5}_{ }$. 
 We employ the shorthand notation 
 $mX\equiv 10^{-X}_{ }$, 
 $pX \equiv 10^{+X}_{ }$.
 The most stringent observational constraint originates from 
 the spectral tilt,
 $n^{ }_\rmi{s} = 0.974 \pm 0.003$~\cite{planck6,act}.
 }
\label{table:sm}
\end{table}
%
%%%%%%%%%%%%%%%%%%%%%%%%%%%%%%%%%%%%%%%%%%%%%%%%%%%%%%%%%%%%%%%%%%%%%

As a second application, we consider a Standard Model embedded
warm inflation scenario, 
recently proposed in ref.~\cite{sm1} (subsequently 
also considered in ref.~\cite{sm2}). The inflaton potential and 
the friction coefficient, which we adopt at face value
from ref.~\cite{sm1}, read
\ba
 V(\bar\varphi)
 & \equiv & \lambda\, {\bar\varphi}^{\hspace*{0.3mm}4}_{ } 
 \;, \quad  \mbox{for} \; |\bar\varphi| \; \gg \; \mbox{GeV} 
 \;, \la{V_sm} 
 \\[2mm] 
%%%%%% 
 \Upsilon(\bar\varphi,T)
 & \equiv & 
 \frac{T^2_{ }}{2 f_a^2}
 \,
 \biggl[ \, 
 \frac{1}{\alphas^5 \Nc^5 T} + \frac{2\Nf^{ }}{\Nc^{ } H (\bar\varphi,T) }
 \,\biggr]^{-1}_{ }
 \;, \quad
 \Nc^{ }\; \equiv \; 3
 \;, \quad
 \Nf^{ }\; \equiv \; 5
 \;. \la{Upsilon_sm}
\ea
For the three parameters ($\lambda$, $f^{ }_a$ and $\alphas^{ }$), 
we consider the same values as in ref.~\cite{sm1}, 
as well as small variants of $f^{ }_a$, 
as tabulated in table~\ref{table:sm}. 
The background evolution is started 
with a common field value, 
large enough to cover all sets,  
$\bar\varphi(t^{ }_\rmi{ref}) \equiv \bar\varphi^{ }_\rmi{ref} 
\equiv 7 \, \mpl^{ }$, at a time
\be
 t^{ }_\rmi{ref} 
 \; \equiv \;
 \sqrt{\frac{3}{8\pi}} \frac{\mpl^{ }}{\sqrt{ V( \bar\varphi^{ }_\rmi{ref} )}}
 \;. \la{tref_sm}
\ee
The initial time derivative $\dot{\bar\varphi}(t^{ }_\rmi{ref})$ 
and temperature $T^{ }_\rmi{ref}$ are fixed in analogy with 
\eqs\nr{init} and \nr{T_ref}, respectively.  
The time $t^{ }_1$ at which $k^{ }_*/(a H) = 10^2_{ }$ and we start evolving
the perturbations, has been determined from 
a match to Standard Model thermodynamics after reheating, 
as explained around \eq\nr{t_1}. 

The observables that we compute are the mean 
value of the scalar amplitude, 
\be
 A^{ }_\rmi{s}
 \; \equiv \; 
 \bigl\langle\,
 \P^{ }_\R(k^{ }_*,t) 
 \,\bigr\rangle^{ }_{ k^{ }_* /( a H ) = 10^{-5}_{ }} 
 \;, \la{As}
\ee
as well as other 
quantities that can be derived from it, 
notably the spectral tilt, 
\be
 n^{ }_\rmi{s} -1
 \; \equiv \; 
 \frac{ {\rm d}\ln \langle\P^{ }_{\R}(k^{ }_*,t)\rangle }{ {\rm d}\ln k } 
 \biggr|^{ }_{k^{ }_* / (a H) = 10^{-5}_{ }}
 \; 
 \approx
 \; 
 \frac{
 \log^{ }_{10} \langle \P^{ }_{\R}( 10^{+\delta}_{ } \, k^{ }_*,t) \rangle
 - 
 \log^{ }_{10} \langle \P^{ }_{\R}( 10^{-\delta}_{ } \, k^{ }_*,t) \rangle
 }{2\delta}
 \; 
 \;. \la{ns}
\ee
The subtraction results in significance loss, 
implying that $\delta$
cannot be taken arbitrarily small;  
a value $\delta\sim 0.1$ is large enough 
for clear variation, but not so large
that $\rmO(\delta^2_{ })$ matters. 
In addition to $A^{ }_\rmi{s}$ and $n^{ }_\rmi{s}$, 
we also evaluate the ratio of tensor and scalar power spectra, 
\be
 r 
 \; \equiv \; 
 \frac{ \P^{ }_\rmi{t} }{  A^{ }_\rmii{s}  }
 \;, \quad
 \P^{ }_\rmi{t}
 \; \approx \; 
 \frac{ 16 H^2_{*} }{ \pi\hspace*{0.3mm} \mpl^2 } 
 \;. \la{r}
\ee
The tensor spectrum gets 
a contribution from thermal noise, 
but it grows as $k^3_{ }$~\cite{sorbo,klose,new}, and is 
hence small in the CMB domain of very small $k$. 
Therefore, it is reasonable to employ the vacuum prediction 
for the tensor spectrum, as indicated in \eq\nr{r}. 

In the top panel of table~\ref{table:sm}, we employ the same input 
parameters as in ref.~\cite{sm1}. We then find $A^{ }_\rmi{s}$
somewhat different from the CMB value 
$A^{ }_\rmi{s} \approx 2.10 \times 10^{-9}_{ }$.
One cause for this 
could be the model dependence discussed
in \se\ref{se:scans}
(ref.~\cite{sm1} employed a fit from ref.~\cite{mehrdad}, 
valid for $\Upsilon\sim T^3_{ }$, 
even though \eq\nr{Upsilon_sm} has a different form).

On the other hand, as shown in the bottom panel of
table~\ref{table:sm}, by adjusting $f^{ }_a$ modestly, we can 
tune $A^{ }_\rmi{s}$ to agree with observation. 
The corresponding $n^{ }_\rmi{s}$
is within $\sim 2\sigma$ of current data 
(though it displays some
differences compared with ref.~\cite{sm1}). 
At this confidence level,
we confirm the viability of the model proposed
in ref.~\cite{sm1}. 

%%%%%%%%%%%%%%%%%%%%%%%%%%% SECTION %%%%%%%%%%%%%%%%%%%%%%%%%%%%%%%%%%%%%%
%
\section{Conclusions and outlook}
\la{se:concl}

Given the growing interest in the physics of density perturbations at 
shorter length scales (larger momenta) than is relevant for the CMB, 
we have developed a framework for computing their evolution
across a smooth reheating period. For this, it has been important
to avoid the approximations adopted, 
if only CMB physics is of interest (cf.\ \se\ref{se:eqs}). 
The purpose of the present study has been to implement this framework 
in its full domain of validity, which requires the inclusion of thermal
fluctuations. To incorporate quantum aspects, 
we have derived the noise autocorrelator through 
a matching to quantum-statistical physics deep inside the Hubble horizon
(cf.\ \se\ref{se:noise}). We have also
shown that the gauge-invariant stochastic 
equations can be studied numerically 
with standard methods (cf.\ \se\ref{se:numerics}).

In addition to reheating, our framework can also be employed 
in the context of warm inflation. 
Apart from comparing with recent literature, 
whose fit forms have been employed in model building
(cf.\ \fig\ref{fig:fit}(right)), 
we have scrutinized the ``universality'' of power spectra that 
originate from warm inflation. 
To offer a new angle on this topic, we have explored a two-parameter
non-powerlaw family of friction coefficients 
(cf.\ \eq\nr{Upsilon_ansatz}), 
motivated by the functional form 
arising from its microscopic derivation
for sphaleron heating. 
The two parameters give us sufficient freedom to 
meaningfully test the premise of universality. 

Our results show that the model dependence of 
$\langle\P^{ }_\R\rangle$ {\em cannot} be 
hidden by reparametrizing the result in terms of 
$T^{ }_*/H^{ }_*$ and $Q^{ }_* \equiv \Upsilon^{ }_*/(3 H^{ }_*)$  
(cf.\ \fig\ref{fig:repara}(left)).
On the other hand, the introduction 
of $\dot{\bar\varphi}^{ }_*/H^{2}_*$ could offer
possibilities for this 
(cf.\ \fig\ref{fig:repara}(right)).
However, $\dot{\bar\varphi}^{ }_*/H^{2}_*$ appears in a complicated
way in the evolution equations
(cf.\ \eqs\nr{dot_R_k}--\nr{dot_S_T}), 
and the dynamics it affects is subtle, as 
$\R^{ }_\varphi$ decouples in the limit where
the $\varphi$ energy
density is subdominant~\cite{alica}
(during inflation only the milder hierarchy
$\dot{\bar\varphi}^2_{ } \ll e^{ }_r$ can be realized). 
Therefore, 
we have not been able to identify a simple freeze-out parameter
that could hide the model dependence, though we do 
not consider it impossible that such a parameter exists. 

As a second application, we have tested the recent proposal
of ref.~\cite{sm1} of embedding warm inflation within the Standard Model. 
Even if we find small differences, requiring a modest adjustment of 
one of the input parameters (cf.\ table~\ref{table:sm}), and even if 
our spectral tilt $n^{ }_\rmi{s}$ is larger in the weak regime,
the qualitative features of the solutions found in ref.~\cite{sm1} 
can be confirmed. In particular, our $n^{ }_\rmi{s}$ is consistent
with current CMB data~\cite{planck6,act} at the $2\sigma$ level.

To summarize, the tests carried out show that
by adopting a more complete framework than has been the norm 
in the warm inflation context
(avoiding gauge fixing and slow-roll approximations, 
and extrapolating thermal noise self-consistently towards the quantum 
domain $k/a \ge T$), one can eliminate theoretical doubts
without inducing additional numerical cost.
Hopefully, in the future, we can use this framework to 
tackle our original main goal, the study of scalar-induced 
gravitational waves originating from the reheating epoch. 

%%%%%%%%%%%%%%%%%%%%%%%%% SECTION %%%%%%%%%%%%%%%%%%%%%%%%%%%%%%%%%%%%%
%
\section*{Acknowledgements}

We thank Sebastian Zell for helpful discussions
and valuable comments on the manuscript, 
and Rudnei Ramos for illuminating correspondence. 

\newpage

%%%%%%%%%%%%%%%%%%%%%%% APPENDIX %%%%%%%%%%%%%%%%%%%%%%%%%%%%%%%%%%%
%
\appendix
\renewcommand{\thesection}{\Alph{section}}
\renewcommand{\thesubsection}{\Alph{section}.\arabic{subsection}}
\renewcommand{\theequation}{\Alph{section}.\arabic{equation}}

%%%%%%%%%%%%%%%%%%%%%%%%% SECTION %%%%%%%%%%%%%%%%%%%%%
%
\section{Inflaton potential, radiation plasma, 
and background initial conditions}
\la{se:V}

The benchmarks and scans 
of \ses\ref{se:benchmarks} and \ref{se:scans} 
focus on understanding the physics originating
from the damping coefficient~$\Upsilon$ and 
the thermal noise $\varrho$, 
so it is helpful to fix the 
vacuum potential and the initial conditions, 
guaranteeing that not too many parameters vary simultaneously. 
For the potential we choose the ansatz referred to as 
``natural inflation''~\cite{ai}, 
\be
 V(\bar\varphi)
 \;\equiv\;
 m^2 f_a^2\, 
 \biggl[ 1 - \cos\biggl( \frac{\bar\varphi}{f^{ }_a} \biggr) \biggr]
 \;, \la{V}
\ee
with the parameter values  
\be
 f^{ }_a = 1.25 \, \mpl^{ } 
 \;, \quad 
 m = 1.09 \times 10^{-6}_{ }\, \mpl^{ }  
 \;. \la{params}
\ee
For $(\Upsilon/H)(k^{ }_*) \ll 1.0$, these yield a curvature power spectrum
$\P^{ }_\R(k^{ }_*) \approx 2.1 \times 10^{-9}_{ }$, 
in good agreement with CMB data.
The initial conditions are taken as 
\be
 \bar\varphi(t^{ }_\rmii{ref})
 \; = \; 3.5 \,\mpl^{ }
 \;, \quad
 \dot{\bar\varphi}(t^{ }_\rmii{ref})
 \; = \; 
 - \frac{V^{ }_{,\varphi}(3.5 \,\mpl^{ })}
        {3 H^{ }_\rmii{ref} + \Upsilon^{ }_\rmii{ref}}
 \;, \la{init}
\ee
where the initial time and its inverse are defined as
\be
  t^{ }_\rmi{ref} 
  \; \equiv \; 
  \sqrt{\frac{3}{4\pi}} 
  \frac{ m^{ }_\rmiii{pl} }{m\, \bar\varphi(t^{ }_\rmiii{ref})}
 \;, \quad
 H^{ }_\rmi{ref} \; \equiv \; t^{-1}_\rmi{ref}
 \;. 
 \la{t_ref} 
\ee
The value of $ \bar\varphi(t^{ }_\rmii{ref}) $  
implies that we start near the top of $V$, 
and $ \dot{\bar\varphi}(t^{ }_\rmii{ref}) $ ensures that we 
soon approach the slow-roll trajectory. However, 
$H^{ }_\rmi{ref} \neq H(t^{ }_\rmi{ref})$, and therefore 
the solution displays an initial transient. 
Afterwards, a period of 
quasi de Sitter expansion follows, with $\dot{H} \ll H^2_{ }$.  
The initial conditions have been chosen ``conservatively'', 
in the sense that, for large~$\Upsilon$, 
the solution undergoes 
hundreds of $e$-folds, before we reach the time $t = t^{ }_1$,
at which we start evolving the perturbations
(cf.\ \fig\ref{fig:cases_background}(left)).
Horizon exit takes place soon afterwards. 

Turning to the properties of the radiation plasma,  
it is represented by the energy density and pressure
\be
 e^{ }_r 
 \; = \; 
 \frac{ g^{ }_* \pi^2 T^4 }{ 30 }
 \;, \quad
 p^{ }_r 
 \; = \; 
 \frac{ g^{ }_* \pi^2 T^4 }{ 90 }
 \;, \quad
 g^{ }_* = 106.75 
 \;, \la{g_star}
\ee
where the value of $g^{ }_*$ is chosen so that the late-time
dynamics can be matched onto a Standard Model plasma. 
The initial condition for $T$ needs to be fixed and, 
even if the exact choice does not matter, transients can be reduced
by searching for an approximate 
stationary temperature following from \eq\nr{bg_T}, 
\be
 \bigl( e^{ }_r + p^{ }_r \bigr)(T^{ }_\rmi{ref})
 \; 
 = 
 \; 
 \frac{\Upsilon^{ }_\rmii{ref}\, \dot{\bar\varphi}^2_{ }(t^{ }_\rmii{ref}) }
 {3 H^{ }_\rmii{ref}}
 \;, \la{T_ref}
\ee
where $T^{ }_\rmi{ref} \equiv T(t^{ }_\rmi{ref})$
and 
$\Upsilon^{ }_\rmi{ref} \equiv \Upsilon^{ }(T^{ }_\rmi{ref})$.

% \newpage

\small{
%%%%%%%%%%%%%%%%%%%%%%%%%%%%%%%%%%%%%%%%%%%%%%%%%%%%%%%%%%%%%%%%%%%%%%%%%%%
%

}

\end{document}